\documentclass[aps,preprint,groupedaddress,showpacs]{revtex4}

\begin{document}
\title{On Magnetic Compression in Gyrokinetic Field Theory}

\author{Bruce D. Scott}
\email[email: ]{bds@ipp.mpg.de}
\affiliation{Max-Planck-Institut f\"ur Plasmaphysik, 
		Boltzmannstr 2,
                D-85748 Garching, Germany}

\date{\today}

\begin{abstract}
The issue of finite magnetic compressibility in low-beta magnetised
plasmas is considered within the gyrokinetic description.  The gauge
transformation method of Littlejohn is used to obtain a Lagrangian
which contains this effect additionally.  The field theory version
obtains a system model which guarantees exact energetic consistency.
Gyrocenter drifts under this model are considered within a
Chew-Goldberger-Low MHD equilibrium allowing for pressure anisotropy.
The contributions to the current divergence balance, hence the
dynamics, due to the difference between the curvature and grad-B
drifts and to the compressibility are shown to cancel up to
corrections of order beta.  This recovers an earlier result with the
same conclusion within linear theory of kinetic ballooning modes.
\end{abstract}


\maketitle

\def\emskip{\hskip 1 em}
\def\hfb{\hfil\break}
\def\etc{{\it etc.}}
\def\visavis{{\it vis-a-vis}\ }
\def\ie{{\it i.e.}}
\def\eg{{\it e.g.}}
\def\etal{{\it et al}}
\def\ua{u.a.\ }
\def\dh{d.h.\ }
\def\zb{z.B.\ }
\def\bzw{bzw.\ }
\def\usw{usw.\ }

\def\idelta{$i$-delta}


\def\half{ {1\over 2} }
\def\third{ {1\over 3} }
\def\fourth{ {1\over 4} }
\def\tth{ {2\over 3} }
\def\twothirds{ {2\over 3} }
\def\threehalves{ {3\over 2} }
\def\fivehalves{ {5\over 2} }
\def\fivethirds{ {5\over 3} }
\def\sevenhalves{ {7\over 2} }
\def\threeh{\threehalves}
\def\eps{\epsilon}
 
\def\grapprox{\mathop{\lower.5ex \hbox{$\buildrel{\fivesy >}\over{\fivesy\sim}$}} \nolimits}
\def\lsapprox{\mathop{\lower.5ex \hbox{$\buildrel{\fivesy <}\over{\fivesy\sim}$}} \nolimits}
\def\grls{\mathop{\lower.5ex \hbox{$\buildrel{\fivesy >}\over{\fivesy <}$}} \nolimits}

\def\vec#1{{\bf #1}}
\def\tsr#1{{\secfnt #1}}
\def\avg#1{\left\langle #1 \right\rangle}
\def\abs#1{\left\vert #1 \right\vert}
\def\prf#1{\overline{#1}}

\def\max{{}_{{\rm max}}}
\def\min{{}_{{\rm min}}}

\def\minus{\mathop{\hbox{--}}\nolimits}

\def\re{\mathop{\rm Re}\nolimits}
\def\im{\mathop{\rm Im}\nolimits}
\def\csch{\mathop{\rm csch}\nolimits}
\def\sech{\mathop{\rm sech}\nolimits}
\def\diag{\mathop{\rm diag}\nolimits}
\def\Max{\mathop{\rm Max}\nolimits}
\def\Min{\mathop{\rm Min}\nolimits}
\def\nint{\mathop{\rm NINT}\nolimits}
\def\mod{\mathop{\rm mod}\nolimits}
\def\det{\mathop{\rm det}\nolimits}
\def\Tr{\mathop{\rm Tr}\nolimits}
\def\sign{\mathop{\rm sign}\nolimits}

\def\LBR{\left\lbrace}
\def\RBR{\right\rbrace}
\def\LB{\left\lbrack}
\def\RB{\right\rbrack}
\def\LP{\left (}
\def\RP{\right )}
\def\qqquad{\qquad\qquad}
\def\qqqquad{\qquad\qquad\qquad}
\def\Det#1{\left\vert\matrix{#1}\right\vert}

\def\pt{\partial}

\def\pzz#1{{\partial #1\over\partial z}}
\def\pxx#1{{\partial #1\over\partial x}}
\def\pyy#1{{\partial #1\over\partial y}}
\def\pww#1{{\partial #1\over\partial w}}
\def\pss#1{{\partial #1\over\partial s}}
\def\prr#1{{\partial #1\over\partial r}}
\def\prhrh#1{{\partial #1\over\partial \rho}}
\def\pthth#1{{\partial #1\over\partial \theta}}
\def\pchch#1{{\partial #1\over\partial \chi}}
\def\ppsps#1{{\partial #1\over\partial \psi}}
\def\pzeze#1{{\partial #1\over\partial \zeta}}
\def\pmumu#1{{\partial #1\over\partial \mu}}
\def\pphph#1{{\partial #1\over\partial \phi}}
\def\ptt#1{{\partial #1\over\partial t}}
\def\pVV#1{{\partial #1\over\partial V}}
\def\phh#1{{\partial #1\over\partial \theta}}
\def\pnn#1{{\partial #1\over\partial \eta}}
\def\pvhvh#1{{\partial #1\over\partial \vartheta}}
\def\pxixi#1{{\partial #1\over\partial \xi}}
\def\dtt#1{{d #1\over dt}}
\def\dss#1{{d #1\over ds}}
\def\drr#1{{d #1\over dr}}
\def\pprr#1{{\partial^2 #1\over\partial r^2}}
\def\pprhrh#1{{\partial^2 #1\over\partial \rho^2}}
\def\ppss#1{{\partial^2 #1\over\partial s^2}}
\def\ppxx#1{{\partial^2 #1\over\partial x^2}}
\def\ppxy#1{{\partial^2 #1\over\partial x\partial y}}
\def\ppxs#1{{\partial^2 #1\over\partial x\partial s}}
\def\ppys#1{{\partial^2 #1\over\partial y\partial s}}
\def\ppyy#1{{\partial^2 #1\over\partial y^2}}
\def\ppzz#1{{\partial^2 #1\over\partial z^2}}
\def\ppww#1{{\partial^2 #1\over\partial w^2}}
\def\pptt#1{{\partial^2 #1\over\partial t^2}}
\def\ppVV#1{{\partial^2 #1\over\partial V^2}}
\def\ppphph#1{{\partial^2 #1\over\partial \phi^2}}
\def\ppthth#1{{\partial^2 #1\over\partial \theta^2}}
\def\pphh#1{{\partial^2 #1\over\partial \theta^2}}
\def\ppnn#1{{\partial^2 #1\over\partial \eta^2}}
\def\ppvhvh#1{{\partial^2 #1\over\partial \vartheta^2}}
\def\ppxixi#1{{\partial^2 #1\over\partial \xi^2}}
\def\ppzeze#1{{\partial^2 #1\over\partial \zeta^2}}
\def\pphze#1{{\partial^2 #1\over\partial\theta\partial\zeta}}
\def\ppz#1{\partial #1/\partial z}
\def\ppx#1{\partial #1/\partial x}
\def\ppy#1{\partial #1/\partial y}
\def\ppw#1{\partial #1/\partial w}
\def\ppr#1{\partial #1/\partial r}
\def\pprh#1{\partial #1/\partial \rho}
\def\pps#1{\partial #1/\partial s}
\def\ppt#1{\partial #1/\partial t}
\def\ppV#1{\partial #1/\partial V}
\def\pph#1{\partial #1/\partial \theta}
\def\ppn#1{\partial #1/\partial \eta}
\def\ppvh#1{\partial #1/\partial \vartheta}
\def\ppxi#1{\partial #1/\partial \xi}
\def\ppze#1{\partial #1/\partial \zeta}

\def\ddt#1{d #1/dt}
\def\pppz#1{\partial^2 #1/\partial z^2}
\def\pppx#1{\partial^2 #1/\partial x^2}
\def\pppy#1{\partial^2 #1/\partial y^2}
\def\pppr#1{\partial^2 #1/\partial r^2}
\def\ppprh#1{\partial^2 #1/\partial \rho^2}
\def\ppps#1{\partial^2 #1/\partial s^2}
\def\pppt#1{\partial^2 #1/\partial t^2}
\def\pppV#1{\partial^2 #1/\partial V^2}
\def\ppph#1{\partial^2 #1/\partial \theta^2}
\def\pppvh#1{\partial^2 #1/\partial \vartheta^2}
\def\pppxi#1{\partial^2 #1/\partial \xi^2}
\def\dddt#1{d^2 #1/dt^2}

\def\grad{\nabla}
\def\cross{{\bf \times}}
\def\div{\grad\cdot}
\def\divp{\grad_\perp\cdot}
\def\divpl{\grad_\parallel\cdot}
\def\curl{\grad\cross}
\def\dpl{\grad_\parallel}
\def\ddpl{\grad_\parallel^2}
\def\dpp{\grad_\perp}
\def\ddpp{\grad_\perp^2}
\def\delsq{\grad^2}
\def\delamb{ \mathchar"0274\hskip -.665em\mathchar"0275 }
\let\delam=\delamb
\def\lapl{\grad^2}
\def\lapldef{\ddpp=(\pt^2/\pt x^2)+K^2(\pt^2/\pt y^2)}

\def\pwww#1{{\partial #1\over\partial \vec w}}
\def\pwwpl#1{{\partial #1\over\partial w_\parallel}}
 
\def\pvv#1#2{{\partial #2\over\partial v_{#1}}}
\def\ppv#1#2{{\partial #2/\partial v_{#1}}}
\def\pvvv#1{{\partial #1\over\partial \vec v}}
\def\pvvp#1#2{{\partial #2\over\partial v'_{#1}}}
\def\ppvp#1#2{{\partial #2/\partial v'_{#1}}}
\def\pvvvp#1{{\partial #1\over\partial \vec v'}}
\def\pvvpl#1{{\partial #1\over\partial v_\parallel}}
 
\def\xunit{\vec{\hat x}}
\def\yunit{\vec{\hat y}}
\def\zunit{\vec{\hat z}}
\def\sunit{\vec{\hat s}}
\def\bunit{\vec{b}}
\def\eunit{\vec{\hat e}}
\def\nunit{\vec{\hat n}}
\def\dt{\Delta t}
\def\becomes{\leftarrow}
\def\from{\leftarrow}
\def\to{\rightarrow}
\def\fromto{\leftrightarrow}
\def\implies{\,\,\,\Longrightarrow\,\,\,}
\def\dotdot{\!:\!}

\def\meters{\,{\rm m}}
\def\invm{\,{\rm m}^{-3}}
\def\invmeter{\,{\rm m}^{-1}}
\def\invsec{\,{\rm sec}^{-1}}
\def\cm{\,{\rm cm}}
\def\km{\,{\rm km}}
\def\invcc{\,{\rm cm}^{-3}}
\def\invcm{\,{\rm cm}^{-1}}
\def\invmm{\,{\rm mm}^{-1}}
\def\mm{\,{\rm mm}}
\def\Vcm{\,{\rm V/cm}}
\def\Acm{\,{\rm A/cm^2}}
\def\kA{\,{\rm kA}}
\def\MA{\,{\rm MA}}
\def\degk{\,{\rm K}}
\def\ergs{\,{\rm erg}}
\def\eV{\,{\rm eV}}
\def\keV{\,{\rm keV}}
\def\kVm{\,{\rm keV/m}}
\def\keVm{\,{\rm keV/m}}
\def\MeV{\,{\rm MeV}}
\def\GeV{\,{\rm GeV}}
\def\kG{\,{\rm kG}}
\def\tesla{\,{\rm T}}
\def\kW{\,{\rm kW}}
\def\MW{\,{\rm MW}}
\def\MWsqm{\,{\rm MW/m^2}}
\def\Wsqm{\,{\rm W/m^2}}
\def\radsec{\,{\rm rad/sec}}
\def\Hz{\,{\rm Hz}}
\def\kHz{\,{\rm kHz}}
\def\MHz{\,{\rm MHz}}
\def\mpersec{\,{\rm m}/{\rm sec}}
\def\msqsec{\,{\rm m^2}/{\rm sec}}
\def\cmsec{\,{\rm cm}/{\rm sec}}
\def\kmsec{\,{\rm km}/{\rm sec}}
\def\mmsec{\,{\rm m}^2/{\rm sec}}
\def\msqsec{\,{\rm m}^2/{\rm sec}}
\def\cmcmsec{\,{\rm cm}^2/{\rm sec}}
\def\ccpersec{\,{\rm cm}^3/{\rm sec}}
\def\minutes{\,{\rm min}}
\def\yr{\,{\rm yr}}
\def\hr{\,{\rm hr}}
\def\Bar{\,{\rm bar}}
\def\sec{\,{\rm sec}}
\def\msec{\,{\rm msec}}
\def\usec{\,\mu{\rm sec}}

\def\ee{\vec E}
\def\bb{\vec B}
\def\ff{\vec F}
\def\jj{\vec J}
\def\qq{\vec q}
\def\aa{\vec A}
\def\kk{\vec k}
\def\vv{\vec v}
\def\uu{\vec u}
\def\xx{\vec x}
\def\ww{\vec w}

\def\bdel{\vec b\cdot\grad}
\def\Bdel{\vec B\cdot\grad}
\def\Jdel{\vec J\cdot\grad}
\def\bdot{\vec B\cdot}
\def\Bdot{\vec B\cdot}
\def\kdot{\vec k\cdot}
\def\exb{\vec E\cross\vec B}
\def\jxb{\vec J\cross\vec B}
\def\uxb{\vec u\cross\vec B}
\def\vxb{\vec v\cross\vec B}
\def\wxb{\vec w\cross\vec B}
\def\ucxb{{\vec u\over c}\cross\vec B}
\def\vcxb{{\vec v\over c}\cross\vec B}
\def\wcxb{{\vec w\over c}\cross\vec B}
\def\jcxb{{\vec J\cross\vec B\over c}}

\def\vexb{\vec v_E}
\def\vpol{\vec v_p}
\def\upol{\vec u_p}
\def\vstar{\vec v_*}
\def\ustar{\vec u_*}
\def\Jstar{\vec J_*}
\def\Jpol{\vec J_p}
\def\vgradb{\vec v_{\grad B}}
\def\qpol{\vec q_p}
\def\qstar{\vec q_\wedge}
\def\qestar{\vec q_e{}_\wedge}
\def\qistar{\vec q_i{}_\wedge}
\def\pistar{\vec\Pi_*}
\def\vR{\vec v_R}
\def\vdl{\vec v\cdot\grad}
\def\vdel{\vec v\cdot\grad}
\def\vedl{\vexb\cdot\grad}
\def\udl{\vec u\cdot\grad}
\def\udel{\vec u\cdot\grad}
\def\uidl{\vec u_i\cdot\grad}
\def\uidel{\vec u_i\cdot\grad}
\def\wdel{\vec w\cdot\grad}
\def\dedt#1{d_E #1/dt}
\def\dett#1{{d_E #1\over dt}}
\def\jpp{J_\perp}
\def\jperp{\vec\jpp}
\def\qpp{q_\perp}
\def\qperp{\vec\qpp}
\def\upp{u_\perp}
\def\uperp{\vec\upp}
\def\wpl{w_\parallel}
\def\wpp{w_\perp}
\def\wperp{\vec\wpp}
\def\vpp{v_\perp}
\def\vperp{\vec\vpp}
\def\lnb{\log B}
 
\def\rms{_{rms}}
 
\def\Jpl{J_\parallel}
\def\jpl{J_\parallel}
\def\Jpp{J_\perp}
\def\jpp{J_\perp}
\def\Jperp{\vec\Jpp}
\def\Bperp{\vec B_\perp}
\def\Apl{A_\parallel}
\def\apl{A_\parallel}
\def\App{A_\perp}
\def\app{A_\perp}
\def\Aperp{\vec\App}
\def\Epl{E_\parallel}
\def\epl{E_\parallel}
\def\Epp{E_\perp}
\def\epp{E_\perp}
\def\Eperp{\vec\Epp}
\def\upl{u_\parallel}
\def\vpl{v_\parallel}
\def\Upl{U_\parallel}
\def\vor{\grad_\perp^2\phi}
\def\kpl{k_\parallel}
\def\kkpl{k_\parallel^2}
\def\kpp{k_\perp}
\def\kperp{\vec\kpp}
\def\kkpp{k_\perp^2}
\def\xpl{{x_\parallel}}
\def\xpp{x_\perp}
\def\DD{\Delta_D}
\def\Dpl{D_\parallel}
\def\Dpp{\Delta_\perp}
\def\Depl{D_e{}_\parallel}
\def\Dipl{D_i{}_\parallel}
\def\Rpl{R_\parallel}
\def\qpl{q_\parallel}
\def\qepl{q_e{}_\parallel}
\def\qipl{q_i{}_\parallel}
\def\Pipl{\Pi_\parallel}
\def\qeperp{\vec q_e{}_\perp}
\def\qiperp{\vec q_i{}_\perp}
\def\mupl{\mu_\parallel}
\def\mupp{\mu_\perp}
\def\nuei{\nu_{ei}}
\def\nuee{\nu_{ee}}
\def\nuii{\nu_{ii}}
\def\wpe{\omega_{pe}}
\def\wpi{\omega_{pi}}
\def\nudamp{\nu_d}
\def\zeff{Z_{\!e\!f\!f}}
\def\lmfp{\lambda_{\!m\!f\!p}}
\def\ws{{\omega_*}}
\def\wsi{{\omega_{*i}}}
\def\wn{\omega_n}
\def\wt{\omega_t}
\def\wi{\omega_i}
\def\wT{\omega_T}
\def\wp{\omega_p}
\def\wc{{\omega_c}}
\def\kappacv{{\cal K}}
\def\wcv{{\omega_B}}
\def\etai{\eta_i}
\def\taui{\tau_i}
\def\rs{\rho_s}
\def\ld{\lambda_D}
\def\Lpl{L_\parallel}
\def\Lpp{L_\perp}
\def\lcorpl{\lambda_\parallel}
\def\lcorpp{\lambda_\perp}
\def\lcorx{\lambda_x}
\def\lcory{\lambda_y}
\def\rch{\rho_{ch}}
\def\npl{\eta_\parallel}
\def\etapl{\eta_\parallel}
\def\ald{a_L}
\def\alde{a_{Le}}
\def\aldi{a_{Li}}
\def\npp{\eta_\perp}
\def\etapp{\eta_\perp}
\def\kappapl{\kappa_\parallel}
\def\dprime{\Delta'}
\def\sk{{}_{\vec k}}
\def\sky{{}_{k_y}}
\def\gk{\gamma_k}
\def\vk{\vfl_k}
\def\nk{\nfl_k}
\def\tk{\tfl_k}
\def\dk{\Delta k}
\def\gd{\gamma_0}
\def\mwn{\Delta_n}
\def\mwh{\Delta_h}
\def\gamT{\Gamma_T}
\def\gamn{\Gamma_n}
\def\gamt{\Gamma_t}
\def\gami{\Gamma_i}
\def\gamc{\Gamma_c}
\def\gamk{\Gamma_k}
\def\gams{\Gamma_s}
\def\gaml{\Gamma_l}
\def\gamr{\Gamma_r}
 
\def\ptb{\widetilde}
\def\psifl{\widetilde\psi}
\def\phifl{\widetilde\phi}
\def\ffl{\widetilde f}
\def\fe{f_e}
\def\fefl{\widetilde f_e}
\def\fifl{\widetilde f_i}
\def\nfl{\widetilde n}
\def\hfl{\widetilde h}
\def\tfl{\widetilde T}
\def\nefl{\widetilde n_e}
\def\nifl{\widetilde n_i}
\def\tefl{\widetilde T_e}
\def\tifl{\widetilde T_i}
\def\pfl{\widetilde p}
\def\pefl{\widetilde p_e}
\def\pifl{\widetilde p_i}
\def\hefl{\widetilde h_e}
\def\vx{\widetilde v_x}
\def\vfl{\widetilde v}
\def\vefl{\widetilde \vexb}
\def\vxfl{\widetilde v_x}
\def\vyfl{\widetilde v_y}
\def\vrfl{\widetilde v_r}
\def\vppfl{\widetilde v_\perp}
\def\vflpp{\widetilde v_\perp}
\def\vplfl{\widetilde \vpl}
\def\Bfl{\widetilde \vec B}
\def\Bflpp{\widetilde B_\perp}
\def\Aplfl{\widetilde A_\parallel}
\def\Appfl{\widetilde A_\perp}
\def\Aperpfl{\widetilde {\vec A}_\perp}
\def\ufl{\widetilde u_\parallel}
\def\vorfl{\grad_\perp^2\phifl}
\def\jfl{\widetilde J_\parallel}
\def\qfl{\widetilde q_\parallel}
\def\qefl{\widetilde q_e{}_\parallel}
\def\qifl{\widetilde q_i{}_\parallel}
\def\jppfl{\widetilde J_\perp}
\def\jperpfl{\widetilde {\vec J}_\perp}
\def\Afl{\ptb A_\parallel}
\def\Jfl{\ptb J_\parallel}
\def\efl{\widetilde E_\parallel}
\def\Efl{\widetilde E_\parallel}
\def\Eppfl{\widetilde E_\perp}
\def\Eperpfl{\widetilde {\vec E}_\perp}
\def\etafl{\widetilde\eta}
\def\isatfl{\widetilde I_{{\rm sat}}}
\def\phiflfl{\widetilde\phi_{{\rm fl}}}
 
\def\teplfl{\widetilde T_e{}_\parallel}
\def\teppfl{\widetilde T_e{}_\perp}
\def\qeplfl{\widetilde q_e{}_\parallel}
\def\qeppfl{\widetilde q_e{}_\perp}
\def\tiplfl{\widetilde T_i{}_\parallel}
\def\tippfl{\widetilde T_i{}_\perp}
\def\qiplfl{\widetilde q_i{}_\parallel}
\def\qippfl{\widetilde q_i{}_\perp}

\def\tepl{ T_e{}_\parallel}
\def\tepp{ T_e{}_\perp}
\def\qepl{ q_e{}_\parallel}
\def\qepp{ q_e{}_\perp}
\def\tipl{ T_i{}_\parallel}
\def\tipp{ T_i{}_\perp}
\def\qipl{ q_i{}_\parallel}
\def\qipp{ q_i{}_\perp}

\def\peplfl{\widetilde p_e{}_\parallel}
\def\peppfl{\widetilde p_e{}_\perp}
\def\piplfl{\widetilde p_i{}_\parallel}
\def\pippfl{\widetilde p_i{}_\perp}

\def\pepl{ p_e{}_\parallel}
\def\pepp{ p_e{}_\perp}
\def\pipl{ p_i{}_\parallel}
\def\pipp{ p_i{}_\perp}


\def\phinn{ {e\phifl\over T} }
\def\nnn{ {\nfl\over n} }
\def\tnn{ {\tfl\over T} }
\def\unn{ {\ufl\over c_s} }
\def\vornn{ \rho_s^2\ddpp\phinn }
\def\jnn{ {\jfl\over ne} }
\def\qnn{ {\qfl\over nT} }
\def\psinn{ {\psifl\over B\rho_s} }

\def\ahat{\hat\alpha}
\def\ehat{\hat\eta}
\def\khat{\hat\kappa}
\def\shat{\hat s}
\def\bhat{\hat\beta}
\def\muhat{\hat\mu}
\def\epss{\hat\epsilon}
\def\bigpoint#1{
    \par\bigskip
    {\baselineskip=\normalbaselineskip
    \parindent=0 pt
    {\hfill\vbox{ #1  }\hfill}}
    \par\bigskip
    }
 
\def\jfm#1{{\it J. Fluid. Mech.} {\secfnt #1}}
\def\jgr#1{{\it J. Geophys. Res.} {\secfnt #1}}
\def\prl#1{{\it Phys. Rev. Lett.} {\secfnt #1}}
\def\physletta#1{{\it Phys. Lett. A} {\secfnt #1}}
\def\pla#1{{\it Phys. Lett. A} {\secfnt #1}}
\def\physlettb#1{{\it Phys. Lett. B} {\secfnt #1}}
\def\pf#1{{\it Phys. Fluids} {\secfnt #1}}
\def\pfa#1{{\it Phys. Fluids A} {\secfnt #1}}
\def\pfb#1{{\it Phys. Fluids B} {\secfnt #1}}
\def\physp#1{{\it Phys. Plasmas} {\secfnt #1}}
\def\nf#1{{\it Nucl. Fusion} {\secfnt #1}}
\def\njp#1{{\it New J. Phys.} {\secfnt #1}}
\def\cpp#1{{\it Contrib. Plasma Phys.} {\secfnt #1}}
\def\cpc#1{{\it Comput. Phys. Comm} {\secfnt #1}}
\def\csd#1{{\it Comput. Sci. Discov.} {\secfnt #1}}
\def\ppcf#1{{\it Plasma Phys. Contr. Fusion} {\secfnt #1}}
\def\jpsj#1{{\it J. Phys. Soc. Japan} {\secfnt #1}}
\def\jpfr#1{{\it J. Plasma Fusion Res. (Japan)} {\secfnt #1}}
\def\jpfrs#1{{\it J. Plasma Fusion Res. SERIES (Japan)} {\secfnt #1}}
\def\plasphys#1{{\it Plasma Phys.} {\secfnt #1}}
\def\revpp#1{{\it Rev. Plasma Phys.} {\secfnt #1}}
\def\iaea#1#2{in {\it Plasma Physics and Controlled Nuclear Fusion
    Research #1}, Proceedings of the #2th International Conference}
\def\EPS#1#2#3{in {\it Proceedings of the
{#1}th European Conference on Controlled Fusion and Plasma Physics,
{#2}, {#3}} (European Physical Society, {#2}, {#3})}
\def\jcp#1{{\it J. Comput. Phys.} {\secfnt #1}}
\def\jetp#1{{\it Sov. Phys. JETP} {\secfnt #1}}
\def\sovjpp#1{{\it Sov. J. Plasma Phys.} {\secfnt #1}}
\def\jnm#1{{\it J. Nucl. Mat.} {\secfnt #1}}
\def\rsi#1{{\it Rev. Sci. Inst.} {\secfnt #1}}
\def\adv#1{{\it Adv. Phys.} {\secfnt #1}}
\def\apjl#1{{\it Astrophys. J. Lett.} {\secfnt #1}}
\def\apj#1{{\it Astrophys. J.} {\secfnt #1}}
\def\astrap#1{{\it Astron. Astrophys.} {\secfnt #1}}
\def\mnras#1{{\it MNRAS} {\secfnt #1}}
\def\vol#1{\ {\secfnt #1}}

\def\dx{\vec {dx}}
\def\dR{\vec {dR}}
\def\dr{\vec {dr}}
\def\dzperp{\vec{dz_\perp}}
\def\drr{\vec r}
\def\de{\vec {de}}
\def\dee{\vec e}
\def\dw{\vec {dw}}
\def\dww{\vec w}
\def\dtheta{d\vartheta}

\def\xdot{\vec {\dot x}}
\def\pdot{\vec {\dot p}}
\def\zzdot{\vec {\dot z}}
\def\Rdot{\vec{\dot R}}
\def\zdot{\dot z}
\def\thetadot{\dot\vartheta}
\def\zpdot{\vec{\dot Z_p}}
\def\pzzp#1{{\pt #1\over\pt\vec Z_p}}
\def\pvpvp#1{{\pt #1\over\pt\varphi}}
\def\ppzpz#1{{\pt #1\over\pt p_z}}
\def\diffa{{\vec d_a}}
\def\diffb{{\vec d_b}}
\def\subR{{}_{\vec R}}
\def\avgR#1{\avg{#1}\subR}

\def\Epsilon{{\cal E}}
\def\Epslash{{\cal E}\hskip -0.2 cm / \hskip 0.05 cm}
\def\epslash{\epsilon\hskip -0.15 cm / \hskip 0.001 cm}

\def\sumsp{\sum_{{\rm sp}}}
\def\sumions{\sum_{{\rm ions}}}
\def\dL{d\Lambda}
\def\dW{d{\cal W}}
\def\dV{d{\cal V}}
\def\intL{\int\dL\,}
\def\intW{\int\dW\,}
\def\intV{\int\dV\,}
\def\scriptl{{\cal L}}
\def\scripte{{\cal E}}
\def\scriptp{{\cal P}}
\def\scriptq{{\cal Q}}
\def\pmumu#1{{\pt #1\over\pt\mu}}
\def\magnet{{\cal M}}

\def\tto{\quad{}\to{}\quad}

\def\pstar{\vec p^*}
\def\Astar{\vec A^*}
\def\Aphi{A^*_\varphi}
\def\bphi{b_\varphi}
\def\Pphi{P_\varphi}
\def\Bstar{\vec B^*}
\def\Bpl{B_\parallel^*}
\def\Bpp{B_\perp}
\def\chiv{\chi'}
\def\pMM#1{{\pt #1\over\pt M}}
\def\phh#1{{\pt #1\over\pt \vartheta}}
\def\pph#1{\pt #1/\pt \vartheta}
\def\Bfl{\ptb B}

\def\gperp{\vec{g_\perp}}
\def\aapl{a_\parallel}
\def\aaperp{\vec{a_\perp}}
\def\bperp{\vec{B_\perp}}
\def\dbpl{\delta B}
\def\zpl{z_\parallel}
\def\zperp{\vec{z_\perp}}
\def\Upl{U_\parallel}
\def\Uperp{\vec{U_\perp}}
\def\rhom{\rho_m}

\def\Kpl{K_\parallel}
\def\Ppl{P_\parallel}
\def\ppl{p_\parallel}
\def\ppp{p_\perp}

\def\bdot{\bunit\cdot}
\def\Bdot{\vec B\cdot}
\def\bdel{\bdot\grad}
\def\Bdel{\Bdot\grad}

\def\equil{(\hbox{eq})}

\section{Introduction}

There is an important result concerning the role of 
magnetic compressibility in electromagnetic responses and
instabilities in a magnetised plasma at low frequency (compared to
gyrofrequencies and compressional Alfv\'en waves) and at low plasma
beta, which has remained quite obscured in current work on magnetised
plasma turbulence.  Basically, the finte plasma beta in confined
plasmas at high thermal performance leads to two effects.  One is the
difference between the grad-B and curvature drifts in a toroidal
magnetic geometry.  The other is the drift due to finite changes in
the strength of the magnetic field, relative to the unperturbed
background.  These effects were shown to cancel in the dynamics by
considering contributions to the divergence balance of the current, 
within the context of general linear stability theory as
represented by kinetic ballooning modes (Tang \etal, 1980, in 
\cite{Tang80}).  The result has
been found to be of significance in energetic particle dynamics
\cite{Chen16}. 

It is of interest to find to what extent this conclusion persists in
the general nonlinear situation represented by the superset of
nonlinear magnetohydrodynamics (MHD), electromagnetic drift wave
(hence drift Alfv\'en) turbulence, and energetic particle dynamics
which can effect either of these.

Gyrokinetic theory (reviewed in \cite{Brizard07})
only produces energetically consistent nonlinear dynamics
when it set up or can be re-cast as a field theory \cite{momcons}.
This means that the field variables are incorporated into a Lagrangian
for the entire system,
so that rather than a treatment of gyrocenter drift
orbits, it becomes a gauge transformation of the entire particle/field
system starting from the case with Maxwell equations and general
Lorentz forces and ending by means of a small set of assumptions with
the case of gyrokinetic field theory \cite{Sugama00,Brizard00}.  It is
possible to do this within the simpler context of Littlejohn's gauge
transformation \cite{Littlejohn83},
while keeping the fields as dynamical variables within
the system Lagrangian \cite{gyrogauge}.  The physical issues regarding
magnetic compression within this context may be addressed by
reproducing this approach while keeping the perturbed perpendicular
magnetic vector potential.  This is the approach taken herein.

There are two main sets of steps to this work.  First is to obtain the
gyrokinetic Lagrangian for the gyrocenter motion and then the
associated field Lagrangian relevant to this situation.  Second is to
examine the gyrocenter drifts and the compressional field response in
the limit of low beta, which is what produces the relevant
cancellation.  What this leads to is simultaneously to drop the
corrections due to the field compression and then to use the curvature
in place of the grad-B drift in the gyrocenter motion.
Essentially, this is to drop the additional perturbed perpendicular
magnetic vector
potential contribution among the fields and then to ensure the model
for the magnetic field unit vector and the magnetic field strength, as
appears in the Lagrangian, produces the same vector for the curvature
and the logarithmic grad-B.  Then, the cancellation is effectively
ensured without any chance of numerical errors emerging to alter 
it (see Acknowledgments, below).

The point of doing this treatment with gyrokinetic field theory is
two-fold.  First, the result is applicable to general levels of
nonlinearity within the low-beta, magnetised plasma regime, not just
``high-$n$'' or small-scale modes within MHD or microinstability
models.  Second, both the appearance of the magnetic compression in
the Hamiltonian ($\dbpl$ in $H$, Eq.\ \ref{eq:hh2}, below) and the
standard representation of the low-beta MHD force balance ($\dbpl$
versus pressure in Eq.\ \ref{eq:gkcompress}, below) emerge natively
from the theory.  They could have been brought in {\it ad hoc}, as is
customary, but of course in that case the result would carry much less
weight.

\section{Setup of the Field Theory}

We will transform the original Maxwell-Lorentz Lagrangian (Landau and
Lifshitz \cite{LandauLifshitz}, nonrelativistic) under the assumptions
of low frequency, quasineutrality, low beta, and flute mode ordering,
into a gyrocenter Lagrangian in canonical form.  For the gyrocenters
this means
\begin{equation}
  L_p = \pstar\cdot\Rdot + M\thetadot - H
  \label{eq:canonicalform}
\end{equation}
in which $\pstar$ is a canonical momentum containing the vector
potential for a strong background guiding magnetic field, depending
on phase space
coordinates but not on time and not on any dynamical field variables,
$M$ is a coordinate (or proportional to a coordinate) representing
gyromotion perpendicular to the background field, and $H$ is the
Hamiltonian depending on phase space coordinates and dynamical field
variables but not time, $\vec R$ is the spatial gyrocenter coordinate,
and $\vartheta$ is the gyrophase angle.

This $L_p$ is then embedded within phase space together with 
a free field Lagrangian density $\scriptl_f$, depending on the 
dynamical field variables and their derivatives.  The structure for
the system is
\begin{equation}
  L = \sumsp\intL f\,L_p + \intV\scriptl_f
  \label{eq:embedded}
\end{equation}
in which $L_p$ is multiplied by a gyrocenter distribution function $f$
and integrated over the phase space for each charged particle species,
then summed over species, and then added to the spatial domain
integral over the field Lagrangian density.  This $L$ is therefore a
functional of all the variables whose dynamics are under consideration.
The gyrocenter motion is
found by setting the functional derivative of $L$ with respect to each
gyrocenter coordinate 
variable to zero, and then the field responses are found by 
setting the functional derivative of $L$ with respect to each field
variable to zero.  In the Landau Lifshitz language, the free particle
Lagrangian is the part of $L_p$ not including field variable
dependence, the free field Lagrangian is all in $\scriptl_f$, and the
interaction Lagrangian, that part containing combinations of the
gyrocenters and the fields, is all in $H$ itself, and solely.  When
gyrocenter motion is considered, $f$ is a set of delta functions
giving the coordinates for each gyrocenter.  When the field responses
are considered, it is a phase space gyrocenter density giving rise to
charge density and current contributions.

The goal of the gyrocenter gauge transformation is to put the system
into this canonical form as well as filtering out high frequency
dynamical responses which are outside of the assumptions of the low
frequency parameter regime (dynamics below the ion gyrofrequency and
as well below magnetosonic compressional frequencies).  The
consideration of the 
gyrocenter drifts and the compressional response then follows all the
consequences of those choices.  Exact energetic consistency is
guaranteed \cite{momcons,Sugama00,Brizard00}.

\section{Magnetic Compression in the Gyrocenter Lagrangian}

We first establish what we mean by canonical form and start with the
textbook case in this language.  Then we do the gauge transformation
for the magnetised plasma case.  The procedure is the one given by
Littlejohn \cite{Littlejohn83}, with the field variables treated self
consistently \cite{gyrogauge}.

\subsection{Canonical Form of the Textbook Case}

We start with the standard electromagnetic Lagrangian of Landau and
Lifshitz \cite{LandauLifshitz}, but in the non-relativistic case, for
the particles,
\begin{equation}
  L_p = {m\over 2}\abs{\xdot}^2 + e\aa\cdot\xdot - e\phi
  \label{eq:lagrangian}
\end{equation}
find the canonical momentum and Hamiltonian
\begin{equation}
  \vec p = {dL_p\over d\xdot} = m\xdot+e\aa
  \qquad
  H = \vec p\cdot\xdot - L_p = {1\over 2m}\abs{\vec p-e\aa}^2 + e\phi
  \label{eq:canonmomentum}
\end{equation}
in which $\pstar$ and $\vec p$ are the same because no geometry is
involved.  Then use this to recast $L_p$ in canonical form
\begin{equation}
  L_p = \vec p\cdot\xdot - H \qquad
  \label{eq:maxwell}
\end{equation}
In Eq.\ (\ref{eq:lagrangian}), 
the first term is the free particle piece and the
next two are the interaction piece (non-relativistic case of the
four-vector $J^aA_a$).  The free field piece is then added, with
$L_p$ embedded in particle phase space, which is
Eq.\ (\ref{eq:embedded}) with $f=f(\vec x,\vec p)$ and
\begin{equation}
  \scriptl_f = \half\LP\eps_0 E^2-\mu_0^{-1} B^2\RP
  \label{eq:maxwellfield}
\end{equation}
Then, Eqs.\ (\ref{eq:embedded},\ref{eq:maxwell},\ref{eq:maxwellfield})
describe the system Lagrangian.  Variation of $\vec p,\vec x$ for each
particle and setting the relevant functional derivatives to zero finds
the equations for the particles,
\begin{equation}
  \dtt{\xx} = {1\over m}(\vec p - e\aa) \qquad
  \dtt{\vec p} = {e\over m}\grad\aa\cdot(\vec p-e\aa) - e\grad\phi
\end{equation}
These are just Hamilton's equations for this $L_p$.
Re-defining the parallel velocity
\begin{equation}
  \vec U = \vec U(\vec x,\vec p) = {1\over m}(\vec p - e\aa)
\end{equation}
as a function of the coordinates, we can recover the equations in
their more familiar form
\begin{equation}
  \dtt{\xx} = \vec U \qquad
  m\dtt{\vec U} = e\LP\vec E+\vec U\cross\vec B\RP
\end{equation}
with $\vec E$ and $\vec B$ in terms of $\phi,\aa$ as
\begin{equation}
  \vec E = -\LP\ptt{\aa}+\grad\phi\RP \qquad
  \vec B = \curl\vec A
  \label{eq:fielddefs}
\end{equation}
We have used the vector identity
\begin{equation}
  \vv\cross(\curl\vec A) = (\grad\aa)\cdot\vv - \vdel\aa
  \label{eq:vecidentity}
\end{equation}
to get the Lorentz force in the familiar form.  Hence canonical form
does not actually change the dynamics, but is merely a representation
which can elucidate computation and physical diagnosis.

Variation of the system Lagrangian with respect to $\phi$ and $\aa$
and then recasting in terms of Eq.\ (\ref{eq:fielddefs}) produces
Maxwell's equations for $\vec E$ and $\vec B$ as in the textbook
case. 

\subsection{The Gyrocenter Langrangian}

We recast $L_p$ in terms of the Poincar\'e-Cartan fundamental one-form,
\begin{equation}
  \gamma = L_p\,dt = \vec p\cdot \vec{dq} - H dt
\end{equation}
with canonical momentum $\vec p$ and particle position $\vec q$
identified as $\vec q=\xx$ and $\vec p$ as in
Eq.\ (\ref{eq:canonmomentum}).

In the case of a magnetised plasma, the background magnetic field is
used as an anchor for momentum by splitting $\vec A$ into a background
piece and then fluctuations in the parallel and perpendicular
directions,
\begin{equation}
  \aa \tto \vec A_T = \aa + \aapl\bunit + \aaperp
\end{equation}
where subscript $T$ denotes the total.
From here on, the background magnetic field is described by its
vector potential $\aa$ (with $\bb=\curl\aa$)
unit vector $\bunit$, and strength $B$.  The perturbations are
$\aapl$ and $\aaperp$, or $\vec a=\aapl\bunit+\aaperp$ as a vector,
with the perturbed magnetic field strength $\dbpl$ given in
terms of $\aapl$ and $\aaperp$ in a way to be described later.

The canonical momentum is split into an equilibrium part ($e\aa$) and
a perturbed part due to the motion and also the field variables.  So
$\vec p$ is replaced by $\pstar$ in the notation, since $\vec p$ was a
coordinate but now $\pstar$ includes the geometry of the background
magnetic field.

The perturbed
canonical momentum is also now split into parallel and
perpendicular parts, so that
\begin{equation}
  \pstar = e\aa + \zpl\bunit + \zperp
\end{equation}
and the parallel and perpendicular velocities are given by
\begin{equation}
  m\Upl = \zpl - e\aapl \qquad
  m\Uperp = \zperp - e\aaperp 
  \label{eq:upl}
\end{equation}
These are auxiliary functions, not coordinates. In particular, the
spatial gradient acting in gyrocenter coordinate space is taken
holding the canonical momenta fixed, not the velocities.

The dynamical variables are $(\xx,\zpl,\zperp)$ for each particle, and
$(\phi,\aapl,\aaperp)$ which cover the electric and magnetic fields,
\begin{equation}
  \vec E = -\LP\ptt{\aapl}\bunit+\ptt{\aaperp}+\grad\phi\RP \qquad
  \delta\vec B = \curl\LP \aapl\bunit + \aaperp \RP
\end{equation}
in this language.  Note that $\vec E$ is arranged with the
dynamical field variables entirely.

The one-form in this representation is now
\begin{equation}
  \gamma = \LP e\aa+\zperp+\zpl\bunit\RP\cdot\dx - H\,dt
  \qquad
  H = {m\over 2}\Upl^2
  + {m\over 2}\abs{\Uperp}^2
  - e\phi
  \label{eq:hhorig}
\end{equation}
together with Eq.\ (\ref{eq:upl}),
in terms of these variables.

Up to here there have been no approximations (except for using the
non-relativistic case), but now we assume the strong background field
and 
take the low frequency limit
\begin{equation}
  \aapl\bunit+\aaperp\ll\aa \qquad \omega\ll\Omega=eB/m
\end{equation}
effectively using $\delta=m/e$ as a parameter for expansion.  The
$e\phi$ term is placed at lowest order, $O(1)$, to be able to have the
velocities, including the ExB velocity, all at the same order, namely
$O(\delta)$.  Then, the gyromotion and the finite gyroradius corrections
to $\phi$ emerge at $O(\delta^2)$.
The particle position $\vec x$ is expanded as
\begin{equation}
  \vec x = \vec R + \drr \qquad 
  \vec R=O(1) \qquad \drr=O(\delta)
\end{equation}
with $\vec R$ and $\drr$ ultimately the gyrocenter position and a
correction to be determined by the condition to transform the
perpendicular part of the particle Lagrangian and in the process
maintain canonical form.  
We therefore have the split arranged as
\begin{equation}
  \gamma_0 = e\aa\cdot\dR - H_0\,dt \qquad
  H_0 = e\phi
  \label{eq:hh0}
\end{equation}
at lowest order $O(1)$, for which the solution is
\begin{equation}
  \Rdot_0 \equiv \uu_0 = \vexb \equiv {1\over B}\bunit\cross\grad\phi
  \label{eq:vexb}
\end{equation}
just the ExB velocity, and then
\begin{equation}
  \gamma_1 = e\aa\cdot\dr
  + \LP e\drr\cdot\grad\aa + \zpl\bunit + \zperp\RP \cdot\dR
  - \LP H_1 + \drr\cdot\grad H_0 \RP dt
\end{equation}
\begin{equation}
  H_1 = {1\over 2m}\LP \zpl-e\aapl\RP^2
  + {1\over 2m}\abs{\zperp-e\aaperp}^2
\end{equation}
at first order $O(\delta)$, and then
\begin{equation}
  \gamma_2 = (e\drr\cdot\grad\aa)\cdot\dr - 
  \LP {1\over 2}\drr\drr\dotdot\grad\grad H_0
  + \drr\cdot\grad H_1 \RP dt
\end{equation}
at second order $O(\delta^2)$.

\subsection{first order}

At first order we subtract the gauge term $d(e\aa\cdot\drr)$ from the
one-form to obtain
\begin{equation}
  \gamma'_1 = 
  e \LP \drr\cdot\grad\aa\cdot\dR - \dR\cdot\grad\aa\cdot\drr\RP
  + \LP\zpl\bunit + \zperp\RP\cdot\dR 
  - \LP H_1 + \drr\cdot\grad H_0 \RP dt
\end{equation}
Then, $\drr$ is chosen to eliminate all perpendicular terms contracted
into $\dR$, so that
\begin{equation}
  \gamma'_1 = \zpl\bunit\cdot\dR - H_1'\,dt
  \qquad
  H_1' =  H_1 + \drr\cdot\grad H_0
\end{equation}
with
\begin{equation}
  - e \LP \drr\cdot\grad\aa - \grad\aa\cdot\drr\RP
  = e\LP \drr\cross\curl\aa \RP = \zperp
  \label{eq:drrcurl}
\end{equation}
or
\begin{equation}
  eB\,\drr\cross\bunit = \zperp
\end{equation}
hence
\begin{equation}
  \drr = {1\over eB}\bunit\cross\zperp
  \label{eq:drr}
\end{equation}
using Eq.\ (\ref{eq:vecidentity}) in Eq.\ (\ref{eq:drrcurl}), and then
crossing with $\bunit$ additionally specifying $\bdot\drr=0$.  
The effect on the Hamiltonian occurs through
\begin{equation}
  \drr\cdot\grad H_0 = -\zperp\cdot\vexb
\end{equation}
and then the square is completed on the perpendicular terms in $H_1$
to obtain
\begin{equation}
  H_1' = {1\over 2m}\LP\zpl-e\aapl\RP^2
  + {1\over 2m}\abs{\zperp-e\aaperp-m\vexb}^2
  - {1\over 2m}\abs{m\vexb+e\aaperp}^2 + {e^2\over 2m}\abs{\aaperp}^2
  \label{eq:hh1}
\end{equation}
Note that $\drr$ no longer explicitly appears, and we are done with
first order.

\subsection{second order}

At second order we subtract the gauge term 
$d(\half e\drr\cdot\grad\aa\cdot\drr)$ from the one-form to obtain
\begin{equation}
  \gamma'_2 = \half\zperp\cdot\dr - H_2'\,dt
  \qquad
  H_2' = {1\over 2}\drr\drr\dotdot\grad\grad H_0
  + \drr\cdot\grad H_1
\end{equation}
using both Eqs.\ (\ref{eq:vecidentity}) and (\ref{eq:drr}).

This first term in $\gamma'_2$ is the one accounting for magnetisation
and gyromotion.  At this point we invoke the existence of the first
order drift velocity to split $\zperp$ into
\begin{equation}
  \zperp = e\aaperp + m\vexb + m\dww
  \label{eq:wwdef}
\end{equation}
where $\dww$ is the fast, thermal gyrating velocity.  Due to the
time scale ordering ($\omega\ll\Omega$), 
\begin{equation}
  \dzperp \tto m\,\dw \qquad \dr \tto {1\over\Omega}\bunit\cross\dw
\end{equation}
are replaced at this order.

The local plane
perpendicular to $\bunit$ is spanned by unit vectors $\dee_1$ and
$\dee_2$ such that $\dee_1\cross\dee_2\cdot\bunit=1$.  The gyrophase
angle is reckoned clockwise from $\vartheta=0$ along $\dee_1$.
We then have
\begin{equation}
  \dww = -w\LP \dee_1\sin\vartheta + \dee_2\cos\vartheta \RP
\end{equation}
and can find
\begin{equation}
  \bunit\cross\dww = -w\LP \dee_2\sin\vartheta - \dee_1\cos\vartheta \RP
\end{equation}
\begin{equation}
  \dw = 
  -w\LP \dee_1\cos\vartheta - \dee_2\sin\vartheta \RP\, d\vartheta
  -w\LP \de_1\sin\vartheta + \de_2\cos\vartheta \RP
  +{dw\over w}\dww
\end{equation}
The magnetisation term is then
\begin{equation}
  \half\zperp\cdot\dr = {m w^2\over 2\Omega}
  \LP d\vartheta - \de_1\cdot\dee_2\RP + O(\delta)
\end{equation}
keeping only the fastest process in this effect.
The term with $\de_1$ is for gyrogauge invariance, with the
combination in parentheses invariant under the angular shift
$\vartheta\to\vartheta+\eta$ for $\eta$ spatially variable.
In practical terms it may be neglected since 
$(mw^2/2\Omega)(\grad\dee_1)\cdot\dee_2$ 
is one fector of $m/e$ smaller in $\Astar$
than $\zpl\bunit$, the leading order term inducing charge separation
due to the curvature of the magnetic field lines.

Next comes $H_2'$, through the FLR corrections, singly
in $H_1$ and quadratically in $H_0$.   The $H_1$ piece arises from
the contribution from $\drr\cdot\grad(U^2/2m)$ entering through
$\aaperp$ to produce the compressibility effect.  We examine this
piece first.  All terms with the parallel velocity $\Upl$ drop
since $\drr$ enters through $\bunit\cross\zperp$.  This leaves
\begin{equation}
  {m\over 2}\drr\cdot\grad\abs{\Uperp}^2 = \drr\cdot
  \LP m\grad \Uperp \RP\cdot\Uperp
\end{equation}
Noting that $\grad\Uperp$ enters only
through $\grad\aaperp$, and using Eq.\ (\ref{eq:drr}), we have
\begin{equation}
  {m\over 2}\drr\cdot\grad\abs{\Uperp}^2
  = -{1\over B}\bunit\cross\zperp\cdot
  \LP \grad\aaperp \RP\cdot\Uperp
\end{equation}
On the right hand side,
the combination $\zperp\Uperp$ appears, effectively as a tensor.  It
expands as 
\begin{equation}
  \zperp\Uperp = m\dww\dww + (m\vexb+e\aaperp)\vexb
  + m\dww\vexb + (m\vexb+e\aaperp)\dww
\end{equation}
This combination is to be gyroaveraged (denoted $\avgR{\cdots}$), using
\begin{equation}
  \avgR{\dww\dww} = {w^2\over 2}\gperp \qquad
  \avgR{\dww\cdot\dww} = w^2 \qquad
  \avgR{\dww} = 0
\end{equation}
where $\gperp=\vec g-\bunit\bunit$ is the perpendicular metric
tensor.  The cross terms with one factor
of $\dww$ gyroaverage to zero, leaving
\begin{equation}
  \avgR{\zperp\Uperp} = {mw^2\over 2}\gperp + (m\vexb+e\aaperp)\vexb
\end{equation}
Only the first term is kept, with the corrections no larger
than ExB Mach number squared, which in practical terms is $O(\delta)$
or smaller.
The metric tensor gives
\begin{equation}
  (\bunit\cross\vec g)\dotdot\grad\aa = \bunit\cdot\curl\aa
\end{equation}
as a vector identity, noting that $\bunit\cross\gperp$ is the same as
$\bunit\cross\vec g$.
Therefore, the first expression in $H_2'$ becomes
\begin{equation}
  \drr\cdot\grad H_1 = {mw^2\over 2B}\bunit\cdot\curl\aaperp
\end{equation}
which can be recast as
\begin{equation}
  \drr\cdot\grad H_1 = {m w^2\over 2B}\dbpl
  \label{eq:hh2}
\end{equation}
where we define the perturbation in the
strength of the magnetic field as
\begin{equation}
  \dbpl = \bunit\cdot\curl\aaperp \qquad
  \label{eq:dbpl}
\end{equation}
and it is multiplied by the effective magnetic moment.

The rest of $H_2'$ is the FLR correction
\begin{equation}
  H_L = \half\drr\drr\dotdot\grad\grad H_0
  = \half\drr\drr\dotdot\grad\grad(e\phi)
\end{equation}
which using the same rules as above becomes
\begin{equation}
  H_L = \fourth\rho_L^2\ddpp(e\phi) 
\qquad
  \rho_L = {w\over\Omega}
  \label{eq:hhl}
\end{equation}
with $\rho_L$ the effective gyroradius and $\dpp$ is the perpendicular
component of the gradient, given by its
contraction with $\gperp$.

In total, we have
\begin{equation}
  \gamma = \LP e\aa+\zpl\bunit \RP\cdot\dR + M\,\dtheta - H\,dt
  \label{eq:gammagk}
\end{equation}
so that
\begin{equation}
  L_p = \LP e\aa+\zpl\bunit \RP\cdot\Rdot + M\,\thetadot - H
  \label{eq:llgk}
\end{equation}
is now the gyrocenter Lagrangian.  The gyrocenter Hamiltonian is
\begin{equation}
  H = {m\over 2}\Upl^2 + e\phi_E + M\Omega_E
  \label{eq:hhgk}
\end{equation}
with $\Upl$ defined in Eq.\ (\ref{eq:upl}) and
\begin{equation}
  e\phi_E = e\phi
  - {1\over 2m}\abs{m\vexb+e\aaperp}^2 + {e^2\over 2m}\abs{\aaperp}^2
\end{equation}
\begin{equation}
  M = {m w^2\over 2\Omega}
  \qquad
  \Omega_E = \Omega + {1\over 2B}\ddpp\phi + {e\over m}\dbpl
\end{equation}
The FLR piece $H_L$, defined in
Eq.\ (\ref{eq:hhl}), manifests as the ExB vorticity as the
second term in $\Omega_E$, and which together with the compressional
piece $\dbpl$ enters $\Omega_E$ as a correction to the gyrofrequency.
The generalised potential $\phi_E$ 
includes the inertia piece which gives rise to polarisation 
(the square of $\vexb$)
as well as additional corrections due to $\aaperp$.  If these are
ordered small, then the only contribution in $L_p$ due to $\aaperp$ is
just the $\dbpl$ piece, which is the conventional approximation.

Inspection of $L_p$ yields that $M$ is an invariant, since the rest of
$L_p$ does not involve the gyrophase angle $\vartheta$.  
This $M$ is just $m/e$
times the effective magnetic moment, essentially the magnetic moment
in the rest frame of the ExB velocity.  Once the low frequency
limit has been taken with the perpendicular dynamics, and we have the
gyrocenter Lagrangian in this form, $M$ becomes an invariant and
we are enabled to
switch coordinates from $\zperp$ to $M$ and $\vartheta$ in the
perpendicular velocity space, so that the gyromotion is merely
$M\dtheta$ and $M$ otherwise only appears in $H$.  The consequence for
magnetosonic compressional waves is discussed below.

Note that the split in Eq.\ (\ref{eq:wwdef}) is the only place where
we have formally used the lowest order drift in Eq.\ (\ref{eq:vexb}).
The form in Eq.\ (\ref{eq:hh1}) is arrived at naturally.  Polarisation
due to the square of $\vexb$ arises naturally through the use of
canonical form in Eq.\ (\ref{eq:hhorig}), the starting point.

\section{The System Lagrangian}

The system Lagrangian is built in the form of Eq.\ (\ref{eq:embedded})
by approximating Eq.\ (\ref{eq:maxwellfield}) in the low frequency
regime including quasineutrality.  Since $\rhom/B^2\gg\eps_0$, where
$\rhom$ is the mass density found by integrating $fm$ over velocity
space and summing over species, ion polarisation overcomes the true
space charge effect and the overall charge density is effectively kept
to zero.  This is done by neglecting $\eps_0E^2$ in the field term, as
elucidated in Ref.\ \cite{Sugama00}.  The resulting free field
Lagrangian is
\begin{equation}
  \scriptl_f = -{1\over 2\mu_0}\abs{\curl\vec A_T}^2
  \label{eq:fieldgk}
\end{equation}
just the magnetic field energy piece with all the pieces kept in the
potential,
\begin{equation}
  \vec A_T = \vec A + \vec a \qquad\hbox{with}\quad
  \vec a = \aapl\bunit + \aaperp
  \label{eq:aat}
\end{equation}
The field equations will be first written with this version of
$\scriptl_f$ before restricting to low-beta cases.

The system Lagrangian is then Eq.\ (\ref{eq:embedded}) with $L_p$
defined in Eq.\ (\ref{eq:llgk}), $H$
defined in Eq.\ (\ref{eq:hhgk}), and the fields 
defined in Eq.\ (\ref{eq:fieldgk}).

\subsection{gyrocenter motion}

In the rest of this discussion we replace $\zpl$ and $\Upl$ with $z$
and $U$ for clarity, always recalling that these represent parallel
components.  Note that $M$ and $\vartheta$
have already replaced $\zperp$ as perpendicular velocity coordinates.

The gyrocenter drifts are found by setting the functional derivatives
of the action (time integral of $L$) with respect to the gyrocenter
coordinates $(\vec{R},z,M,\vartheta)$ to zero.  We use the fact that
the Lagrangian has canonical form,
as per Eq.\ (\ref{eq:canonicalform}) and the discussion below it.
The system Lagrangian has the same form as per
Eq.\ (\ref{eq:embedded}), in which the time component is
represented by $H$ and $\scriptl_f$ but not the other pieces.

Under canonical form, the gyrocenter drifts are
\begin{equation}
  \Bpl\Rdot = \pzz{\Astar}\cross\grad H + \pzz{H}\Bstar
  \qquad
  \Bpl\zdot = -\Bstar\cdot\grad H
\end{equation}
\begin{equation}
  \dot M = 0 \qquad \thetadot = \pMM{H} - \pMM{\pstar}\cdot\Rdot
\end{equation}
in which
\begin{equation}
  \pstar=e\Astar \qquad \Bstar=\curl\Astar \qquad
  \Bpl = \pzz{\pstar}\cdot\Bstar
\end{equation}
recalling that $\pph{}$ on any quantity in $L_p$ is zero and that
the curl is taken with $(z,M,\vartheta)$ held constant.
These forms are the general ones, once
given Eqs.\ (\ref{eq:canonicalform},\ref{eq:embedded}).
We will go to a more explicit description
after addressing the field equations and then the low-beta
limits first to $H$ and then the field equations, and then proceed
with the drifts.

\subsection{field equations}

The field equations are found by setting the functional derivatives
of the action (time integral of $L$) with respect to the dynamical 
field variables $(\phi,\aapl,\aaperp)$ to zero.  The
variations of $\aapl$
and $\aaperp$ can be done together in terms of $\vec a$ as the
combined vector, using
\begin{equation}
  \aapl=\bunit\cdot\vec a \qquad \aaperp=\gperp\cdot\vec a
\end{equation}
and noting that $\vec g$ and $\bunit$ are not varied.

Variation due to $\phi$ yields
the polarisation (gyrokinetic Poisson) equation
\begin{equation}
  \sumsp\intW\LB ef - {1\over\Bpl}\div \Bpl{f\over B}\bunit\cross
  \LP m\vexb+e\aaperp \RP \RB = 0
\end{equation}
where the integration is over the velocity space variables
$(z,M,\vartheta)$ and the sum is over species.  Note how $\Bpl$
functions as the velocity space volume element in the divergence.
Also, $\dW/\Bpl$ commutes past the spatial gradient and annihilates
$\ppz{}$ in any integration.  
This reflects the manifestly covariant nature of the
theory.  The $\vexb$ contribution contains $\phi$ with $\phi/B$ the
effective stream 
function (the scales of variation of $\phi$ and $B$ are very
disparate), so this is the twice gradient of $\phi$ piece which
represents polarisation.

Variation due to $\vec a$ collectively yields
the induction (gyrokinetic Amp\`ere) equation
\begin{equation}
  \curl\LP\curl \vec A_T\RP = \mu_0\LP\Jpl\bunit+\Jperp\RP
\end{equation}
with $\vec A_T$ given in Eq.\ (\ref{eq:aat}),
where the parallel and perpendicular inductive currents are
\begin{equation}
  \Jpl = \sumsp\intW eU\,f \qquad
  \Jperp = \sumsp\intW \LB ef\vexb
  - {1\over\Bpl}\gperp\cdot\curl\LP\Bpl{e\over m} Mf\bunit\RP \RB
\end{equation}
namely, the ones coming from $\delta\aapl$ and $\delta\aaperp$,
respectively.  Note that $\vexb$ is already perpendicular, so that
$\gperp\cdot\vexb=\vexb$.
Since the equilibrium piece $\aa$ is in this equation the equilibrium
current has to be in the respective pieces of $\Jpl$ and $\Jperp$
(\ie, $eUf$ contains the equilibrium parallel current and $Mf$ 
contains the equilibrium pressure).  In a linear or delta-f case the
equilibrium with finite pressure and current is given, and this $f$ is
replaced by the disturbance and $\aa$ is deleted from the left hand
side.  Either way, this is the version of the induction equation one
should use in a numerical scheme which requires the magnetic potential
as a vector.

We note that going to the low frequency regime with $M$ and
$\vartheta$ replacing $\zperp$ as perpendicular velocity coordinates
has resulted in the absence of a term in $H$ which is quadratic in
$\aaperp$ (note the two additions to $e\phi$ in $\phi_E$ are written
as pure squares, though the squared amplitude of $\aaperp$ does not
appear when the expressions are expanded).  Therefore, there is no
``skin term'' due to $\aaperp$, which reflects the absence of
magnetosonic compressional waves once the conversion from $\zperp$ to
$(M,\vartheta$) has been made.  More on this below.  For this reason,
there is no $\aaperp$ contribution to the polarisation current piece
due to $fe\vexb$
in $\Jperp$.  Finally, these are magnetisation currents, not the
perpendicular gyrocenter drift currents.  The term due to
$fM$ yields the nominal magnetisation current due to pressure.  The
polarisation piece combines with third order terms to produce the
entirety of the magnetisation current due to polarisation
\cite{Miyato15}.  This will not be pursued here because this piece
will drop anyway when we go to the low-beta regime.

\subsection{the low-beta limit}

This type of model is done under the low-beta approximations, where
beta enters as both
\begin{equation}
  \beta_e = {\mu_0 p_e\over B^2}\qquad\hbox{and}\quad
  \beta = {2\mu_0 p\over B^2}
\end{equation}
where $p_e$ is the electron pressure and $p$ is the total plasma 
pressure.  These are the electron dynamical beta, entering through the
shear Alfv\'en parallel dynamics, and the plasma beta, entering
through pressure induced compression.  The usual situation is 
\begin{equation}
  \beta_e\sim (\Kpl\Lpp)^2 \qquad\hbox{but}\quad \beta\ll 1
\end{equation}
where $\Kpl$ is the parallel wavenumber of the parallel connection
length of the magnetic field lines and $\Lpp$ is the perpendicular
scale of the thermal gradient.  Under these conditions shear Alfv\'en
dynamics is retained by the model but compressional magnetosonic
dynamics is not.  The ordering on these is
\begin{equation}
  \kpp v_A\sim\Omega \qquad\hbox{but}\quad \kpl v_A\sim\omega
\end{equation}
representing 
the usual case in this type of dynamics, where $\kpl$ and $\kpp$
represent parallel and perpendicular wavenumbers, and $\omega$
frequencies, in the spectral range of the dynamics.
Hence $\Kpl\Lpp$ is also
required to be small.  The foregoing collectively produces both the
low frequency and the low beta limit.

The main point is that $e\aaperp$ is small compared to $m\vexb$ at low
beta.  In ExB related turbulence, the ExB vorticity scales as
$c_s/\Lpp$ while the ion inertial drift scale is $\rs=c_s/\Omega$,
with 
\begin{equation}
  c_s^2={T_e\over m_i} \qquad \rs^2={ m_i T_e\over e^2B^2} 
\end{equation}
namely, both defined in terms of the electron
temperature and ion mass.  We also have $\rs c_s=T_e/eB$ as a useful
relation. 
The amplitude of the pressure disturbance is also $O(\delta)$ small
compared to the equilibrium pressure, so that the ratio of the
compressional gyrofrequency correction to the ExB vorticity is
\begin{equation}
  {e\over m}\dbpl \sim {e\over m}\mu_0{p\over B}{\rs\over\Lpp}
  \quad\hbox{to}\quad {1\over B}\ddpp\phi \sim {c_s\over\Lpp}
\end{equation}
which becomes
\begin{equation}
  {e\over m}\dbpl\quad\hbox{to}\quad {1\over B}\ddpp\phi
  \quad\hbox{as}\quad 
  \beta
  \quad\hbox{to}\quad
  1
\end{equation}
that is, $(e/m)\dbpl$ is $O(\beta)$ small compared to the ExB
vorticity.  By the same measure, using the similarity of the length
scales in the dynamics, we have
\begin{equation}
  e\aaperp
  \quad\hbox{to}\quad
  m\vexb
  \quad\hbox{as}\quad 
  \beta
  \quad\hbox{to}\quad
  1
\end{equation}
that is, $e\aaperp$ is $O(\beta)$ small compared to the ExB momentum.
Therefore the term due to $e\aaperp$ is small compared to $m\vexb$
wherever they occur together.  We do allow, however, for the gradient
in $\dbpl$ to somehow matter, so we keep it in $\Omega_E$
while neglecting $e\aaperp$ in the polarisation.  

We essentially have $e\aaperp\ll m\vexb$ by $O(\beta)$
in dynamics not directly driven by magnetic compression (\ie,
compressional responses in ExB dynamics driven by pressure gradients
or currents).
When this is the case, the simplification in the Hamiltonian is
\begin{equation}
  e\phi_E \quad\hbox{becomes}\quad
  e\phi - {m\over 2B^2}\abs{\dpp\phi}^2
\end{equation}
The resulting simplification in the polarisation is
\begin{equation}
  \bunit\cross\LP m\vexb+e\aaperp\RP \quad\hbox{becomes}\quad
  \bunit\cross\LP m\vexb\RP = - {m\over B}\dpp\phi
\end{equation}
namely, the term due to $\aaperp$ drops and the vector operations
involving $\bunit$ produce the inertial polarisation as in the
conventional case without the extra $\vexb$ contribution to the FLR
correction to $e\phi$ \cite{Miyato09,momcons}.
The resulting simplification in the current is to drop the
polarisation term in $\Jperp$, so that it becomes reasonable to split
the Amp\'ere's law into parallel and perpendicular components,
\begin{equation}
  \bunit\cdot\curl\curl\LP\aapl\bunit\RP = \mu_0\Jpl
  \qquad
  \bunit\cdot\curl\aaperp = \dbpl = -{\ppp\bunit\over B}
\end{equation}
where in the second expression we have dropped the outside curl 
and defined
\begin{equation}
  \ppp = \sumsp\intW M\Omega\,f \qquad
\end{equation}
This simplified form of the equation for $\dbpl$ is the
conventional one in small scale disturbances in the MHD force balance,
as noted in a review of energetic particle modes \cite{Chen16}.  It
was also used in the work by Tang \etal\ on
kinetic ballooning modes \cite{Tang80}.

With $\aapl$ and $\aaperp$ separating like this,
we go back to the magnetic energy term in the
system Lagrangian itself and modify it to reflect these simplifications.
The
background piece is not written since it is regarded as time
independent.  The other pieces are kept as quadratic forms so that
\begin{equation}
  \scriptl_f = -{1\over 2\mu_0}\LP \abs{\grad\aapl}^2 + \dbpl^2 \RP
\end{equation}
with $\dbpl$ defined in terms of $\aaperp$ in Eq.\ (\ref{eq:dbpl}).
The particle Lagrangian is simplified into
\begin{equation}
  L_p\,dt = e\Astar\cdot\dR + M\,\dtheta - H\,dt
\end{equation}
with 
\begin{equation}
  e\Astar = e\aa + z\bunit
  \qquad
  H = e\phi_E + {m\over 2}U^2 + M\Omega_E
\end{equation}
\begin{equation}
  e\phi_E = e\phi - {m\over 2B^2}\abs{\dpp\phi}^2
  \qquad
  mU = z - e\aapl
\end{equation}
\begin{equation}
  \Omega_E = \Omega + {1\over 2B}\ddpp\phi + {e\over m}\,\dbpl
\end{equation}
where we note that $\dbpl$ replaces $\aaperp$ which no longer
explicitly appears, and that the ExB vorticity as a correction to the
gyrofrequency is simply the FLR correction to $e\phi$ with
$\rho_L^2/4$ replaced in terms of $M/2$.  The foregoing is the
specification of the dynamical system at low beta.

For clarity we will neglect the FLR term in $\Omega_E$ in deriving and
discussing the resulting quantities as their form and function are not
affected by its presence or absence.

\subsection{low-beta field equations}

The field equations are found once again by setting the functional
derivatives of the action (time integral of $L$) with respect to the
dynamical field variables, now $(\phi,\aapl,\dbpl)$, to zero.

Variation due to $\phi$ yields the new polarisation equation 
\begin{equation}
  \sumsp\intW\LB ef
  + {1\over\Bpl}\div\Bpl{fm\over B^2}\dpp\phi\RB = 0
  \label{eq:gkpol}
\end{equation}
Doing the integration and species sum yields
\begin{equation}
  \div\vec P_G = \rho_G \qquad
  \rho_G = \sumsp\intW ef \qquad
  \vec P_G = - \sumsp\intW {fm\over B^2}\dpp\phi
  = - {\rhom\over B^2}\dpp\phi
  \label{eq:polarisation}
\end{equation}
in terms of $\rho_G$, the gyrocenter charge density, and $\vec P_G$, the
gyrocenter polarisation.  Using $\eps_0\ll\rhom/B^2$,
this equation sets the true charge density to
zero while allowing a finite $\div\vec E$, the meaning of
quasineutrality. 

Variation due to $\aapl$ yields
the new induction equation
\begin{equation}
  \ddpp\aapl + \mu_0\sumsp\intW eU\,f = 0
  \label{eq:gkind}
\end{equation}
The form with the skin term effect is found by splitting $mU=z-e\aapl$
and arranging the $\aapl$ term on the left.
Doing the integration and species sum yields
\begin{equation}
  \ddpl\aapl + \mu_0\Jpl = 0 \qquad
  \Jpl = \sumsp\intW eU\,f
  \label{eq:current}
\end{equation}
the conventional form of the parallel Amp\`ere's law in terms of the
parallel current.  If the scale of variation of $\bunit$ is ordered
large, this is the resulting form.  Factors of toroidal major radius
appear in this equation if using a global tokamak model.

Variation due to $\dbpl$ yields
the new compressional response
\begin{equation}
  \dbpl + {\mu_0\over B}\sumsp\intW M\Omega\,f = 0
\end{equation}
Doing the integration and species sum yields
\begin{equation}
  \dbpl = - {\mu_0\over B}\ppp
  \label{eq:gkcompress}
\end{equation}
recovering the conventional form of the perpendicular force balance
in terms of the perpendicular pressure as noted above 
\cite{Tang80,Chen16}.

This factor of $\ppp$ in Eq.\ (\ref{eq:gkcompress}) reflects either
fluctuations upon an established equilibrium, or the entire
perpendicular pressure, depending on the nature (completeness) of the
model under consideration.  Indeed, the same is true of the parallel
current $\Jpl$.  If the background is an established equilibrium, then
$\Jpl$ represents only fluctuations.  If the background is the
toroidal vacuum field in a tokamak, then both $\ppp$ and $\Jpl$ will
represent the entire quantity.  More on this later when we consider
cancellations in the dynamics.

\subsection{low-beta gyrocenter drifts}

In the Lagrangian we have here, the canonical momentum has its
conventional form,
\begin{equation}
  e\Astar = e\aa+z\bunit \qquad
  \Bstar = \curl\Astar = \vec B + {z\over e}\curl\bunit
\end{equation}
\begin{equation}
  \Bpl = e\pzz{\Astar}\cdot\Bstar = B+{z\over e}\bdot\curl\bunit
\end{equation}
recalling that the curl is taken at constant $z$.  In the
electrostatic case $z=m\vpl$, and then these forms reduce to those in
Ref.\ \cite{Hahm88}.

The derivatives of $H$ are
\begin{equation}
  \grad H = e\grad\phi_E - {eU\over m}\grad\aapl + M\grad\Omega_E
  \qquad
  \pzz{H} = U
\end{equation}
These go into the drifts to produce
\begin{equation}
  \Bpl\Rdot = \bunit\cross\grad
  \phi_E
  + U\vec B_T
  + {1\over e}mU^2\curl\bunit
  + {1\over e}M\Omega\bunit\cross\grad\log B
  + {1\over eB}M\Omega\bunit\cross\grad(\dbpl)
  \label{eq:gkdrifts}
\end{equation}
in the neglect of the FLR term in $\Omega_E$,
where
\begin{equation}
  \vec B_T = \vec B + \curl(\aapl\bunit)
\end{equation}
The pieces respectively
give the generalised ExB drift including polarisation corrections,
the parallel motion in the total magnetic field including $\aapl$,
the curvature drift, the main grad-B drift, and the correction due to
$\dbpl$, which will give the compressional effect.  The curvature
term in $\Rdot$ includes $zU$, but the $\aapl$ piece in $z$ combines
with the $\grad\aapl$ term in $\Rdot$ to produce the form with 
$\vec B_T$ and then with $mU^2$ in the curvature drift, as is
conventional.  Note that $\div\vec B_T$ vanishes exactly and that
$\aaperp$ does not appear since no term with $\Upl\aaperp$ appears in
any version of $H$ we have considered.
The only additional pieces here not in the conventional
drift kinetic form are the corrections to $B$ in $\Bpl$ and the
corrections to $\phi$ in $\phi_E$ due to polarisation.  Effectively,
polarisation is the main difference between drift kinetic and
gyrokinetic, not the presence or absence of the FLR correction, which
in this low-$\kpp$ form is actually common to both models.

This is the conventional result, but we choose to give its derivation
in terms of the gyrokinetic model rather than to invoke it
externally.  It therefore satisfies exact energetic 
consistency.  We get that by
obtaining the gyrocenter drift flow and the field equations from the
same system Lagrangian, as elucidated elsewhere
\cite{momcons,Sugama00,Brizard00}. 

\section{The Current Under Compressibility}

The equation which determines the MHD response to an imbalance of
forces is
\begin{equation}
  \rhom\LP\ptt{}+\vdel\RP\vv = \jxb-\grad p
  \label{eq:mhdforce}
\end{equation}
Under Reduced MHD (\cite{Strauss76,Strauss77}) it can be written as
\begin{equation}
  \ptt{\varpi} + \div(\varpi\vexb)
  = \vec B_T\cdot\grad{\Jpl\over B}
  + \div\LP {\bunit\over B}\cross\grad p\RP
\end{equation}
where
\begin{equation}
  \varpi = \div{\rhom\over B^2}\dpp\phi
  \label{eq:varpi}
\end{equation}
is proportional to the ExB vorticity, following the velocity ordering
in which advection by the ExB velocity is dominant.  It is important
to have this equation as an exact divergence \cite{momcons}.  
Since Reduced MHD was originally
derived from a constant $\bunit$ this consideration
wasn't fully appreciated, but the role of this equation
is in fact the statement that the total divergence of the
current under quasineutrality is zero.  With variable $\bunit$,
it is formed by crossing the
MHD force equation with $\bunit$ and then applying the divergence, and
then some Reduced MHD ordering on the perpendicular scales of motion.

In the gyrofluid case, the difference in the gyrocenter densities is
the polarisation, whose dominant portion is the same $\varpi$ in
Eq.\ (\ref{eq:varpi}), so that each gyrocenter continuity equation
multiplied by its specific charge forms a partial charge continuity
equation, and their sum yields the gyrocenter charge continuity
equation which is equivalent to, has the same form, and has the same
role in the dynamics as the ExB vorticity equation in a Reduced MHD or
two fluid model \cite{aps99,GEM,braggem}.

In the gyrokinetic case, the corresponding equation is found by
finding the gyrocenter current and then setting its divergence equal
to the time derivative of the polarisation \cite{Miyato09,momcons}.
It is the same as taking
the time derivative of Eq.\ (\ref{eq:gkpol}),
\begin{equation}
  \div\LP \ptt{\vec P_G} + \vec J_G \RP = 0
  \label{eq:gkcharge}
\end{equation}
in which $\vec P_G$ is the gyrocenter 
polarisation given by Eq.\ (\ref{eq:polarisation}).
The time derivative of $\rho_G$ in terms of $f$ is given by
\begin{equation}
  e\Bpl\ptt{f} + \div\Bpl ef\Rdot + \pzz{}\Bpl ef\zdot = 0
\end{equation}
in divergence form (note $\dot M=0$ and $\ppvh{}=0$).  Integration
annihilates the $\ppz{}$ term and produces a charge continuity
equation. 
The quantity under the divergence in this equation is the gyrocenter
current, $\vec J_G$, defined by
\begin{equation}
  \vec J_G = \sumsp\intW ef\Rdot
\end{equation}
and we have the drifts in Eq.\ (\ref{eq:gkdrifts}).
The time derivative of $\rho_G$ produces the divergence of the time
derivative of $\vec P_G$, and since the phase space geometry is
independent of time the divergence and time derivative may be
exchanged.  This produces the divergence of an overall current, which
due to quasineutrality (the shadowing of the Maxwell displacement
current by polarisation) is then divergence free.  The result is the
charge conservation equation in Eq.\ (\ref{eq:gkcharge}).

For these purposes it is sufficient to neglect the $O(\delta)$
correction to $B$ in $\Bpl$ for maximum clarity.  Once the main point
is understood, one can go back and compute the relevant corrections,
including also the FLR term.
Here we obtain $\vec J_G$ and then in the next section examine its
role in the dynamics.

After integration over velocity space and summing over species over
the drifts in Eq.\ (\ref{eq:gkdrifts}), we have
\begin{equation}
  \vec J_G = \rho_G{\bunit\over B}\cross\grad\phi_E
  + {\Jpl\over B}\vec B_T
  + {\ppp\over B}\bunit\cross\grad\log B
  + {\Ppl\over B}\curl\vec b
  + {\ppp\over B^2}\bunit\cross\grad(\dbpl)
\end{equation}
where the relevant moment quantities are $\rho_G$ from
Eq.\ (\ref{eq:polarisation}), $\Jpl$ from Eq.\ (\ref{eq:current}), and
the pressure tensor components given by
\begin{equation}
  \ppp=\sumsp\intW M\Omega\,f \qquad
  \Ppl=\sumsp\intW mU^2\,f
\end{equation}
Note that the parallel one, $\Ppl$ is so labelled because it includes
the finite Mach number contribution from parallel flows.  

This is the gyrocenter current arising from the Lagrangian in the
neglect of FLR corrections and the difference between $\Bpl$ and $B$,
and also under the low Mach number and low beta limits.  Given that
\begin{equation}
  \curl\bunit=(\bdot\curl\bunit)\bunit + \bunit\cross(\bdel\bunit)
\end{equation}
this form of $J_G$ corresponds closely to the traditional drift
kinetic result.  If the parallel Mach number is neglected,
then we have $\Ppl\to\ppl$ as in the standard case in the curvature
drift. 

\section{The Current in the Dynamics}

The gyrocenter current enters the dynamics through its finite
divergence.  Since the time dependent polarisation current compensates
for that, its maintenance of the divergence free total current is what
controls the response of the ExB flow, through the proportional nature
of the ExB vorticity to the gyrocenter charge density; equivalently,
the polarisation.  This produces essentially the same vorticity
equation as in Reduced MHD \cite{braggem,Miyato09,momcons}, 
and it has the same role in the dynamics
(cf.\ Eq.\ \ref{eq:gkcharge}).

The issue here is to compute the divergence of $\vec J_G$ and note
before any further analysis the cancellation between the
difference between the curvature and grad-B drifts on the one hand,
and the response due to $\dbpl$, namely, the magnetic 
compression, on the other.

The paper by Tang \etal\ notes the difference in the equilibrium due
to the finite 
pressure gradient \cite{Tang80}.  In equilibrium,
\begin{equation}
  \jxb = \grad p
\end{equation}
\begin{equation}
  \Bdel\vec B = \mu_0\grad p + \half\grad B^2
\end{equation}
\begin{equation}
  \bdel\bunit = {\mu_0\over B^2}\grad p + \dpp\log B
\end{equation}
\begin{equation}
  \bunit\cross\grad\log B
  = \bunit\cross(\bdel\bunit)
  - {\mu_0\over B^2}\bunit\cross\grad p
  \label{eq:mhdeqp}
\end{equation}
the result after use of Amp\`ere's law and separating $\vec B$ into
$\bunit$ and $B$.  Note that $\bunit\cross\dpp$ is $\bunit\cross\grad$.

Since in the gyrokinetic case we have perpendicular and parallel
pressure entering separately, we have the divergence of the
pressure tensor in place of
the pressure gradient.  For this analysis we use the Chew Goldberger
Low (CGL) pressure tensor \cite{CGL56}, so that $\grad p$ becomes a
divergence, 
\begin{equation}
   \grad p \tto \div\Big[\ppp\vec g + (\Delta p)\bunit\bunit\Big]
\end{equation}
where the pressure and its anisotropy are related to $\ppp$ and $\ppl$
by
\begin{equation}
  p = {\ppl+2\ppp\over 3} \qquad \Delta p = \ppl-\ppp
\end{equation}
Putting this pressure tensor into the MHD equilibrium yields
\begin{equation}
  \bunit\cross\grad\log B
  = \bunit\cross(\bdel\bunit) - {\mu_0\over B^2}\bunit\cross\grad\ppp 
  - {\mu_0\Delta p\over B^2}\bunit\cross(\bdel\bunit)
  \label{eq:cgleqp}
\end{equation}
The result, also present in earlier literature 
(cf., \eg, \cite{Chen91}), 
is in clear analogy to the MHD result in
Eq.\ (\ref{eq:mhdeqp}).  The main point of this is that the main
pressure term appears with $\ppp$, the moment over $M\Omega$, not with
$\ppl$ or $p$.  This will be needed in the cancellation.  

When $\div\vec J_G$ is calculated, the divergence in the magnetic
drift terms acts principally on the pressure, since the scale of the
pressure gradient is much shorter than that of the magnetic field
variations.  The grad-B drift term is
\begin{equation}
  \div{\ppp\over B}\bunit\cross\grad\log B
  = \grad{\ppp\over B}\cdot\bunit\cross\grad\log B
  + {\ppp\over B}\div\LP \bunit\cross\grad\log B\RP
\end{equation}
in which the first term on the right hand side expands into
\begin{equation}
  \grad{\ppp\over B}\cdot\bunit\cross\grad\log B
  = \grad{\ppp\over B}\cdot\LB \bunit\cross(\bdel\bunit)
  - {\mu_0\over B^2}\bunit\cross\grad\ppp\equil 
  - {\mu_0\Delta p\over B^2}\bunit\cross(\bdel\bunit)\RB
  \label{eq:gradbpiece}
\end{equation}
Here it is important to note that the equilibrium pressure is the one
which enters the difference between $\bdel\bunit$ and $\grad\log B$,
while $\ppp$ arising from the moment over $M\Omega$ is the entire
dynamical variable.

The contribution due to the compressional response $\dbpl$ is found in
a similar manner and is given by
\begin{equation}
  \div\vec J_G = \cdots + \grad{\ppp\over B^2}\cdot\bunit\cross\grad(\dbpl)
\end{equation}
Substituting $\dbpl$ using its field equation,
Eq.\ (\ref{eq:gkcompress}), gives
\begin{equation}
  \div\vec J_G = \cdots - 
  \grad{\ppp\over B^2}\cdot\bunit\cross\grad{\mu_0(\delta\ppp)\over B}
  \label{eq:deltabpiece}
\end{equation}
where it is noted that the compressional response is given by the
perturbation.  More detail on these distinctions is given shortly.

The main point is that these two sets of terms add, so that the set of
terms containing $\ppp$ is
\begin{equation}
  \div\vec J_G = \cdots 
  + {1\over B}\LB \bunit\cross(\bdel\bunit)\cdot\grad\ppp
  - {\mu_0\over B^2}\grad\ppp
  \cdot\bunit\cross\grad(\ppp\equil+\delta\ppp) \RB
  + O(\beta)
\end{equation}
in which the first term is the main curvature term, the rest of the
square brackets is the combination of the equilibrium curvature
correction to the grad-B drift and the compressional response (the
equilibrium and fluctuating pieces, respectively.  All other
contributions contain derivatives of the magnetic structure instead of
$\grad\ppp$, which scale as the curvature term, and are preceded by a
coefficient of $\mu_0\ppp/B^2$ or $\mu_0(\Delta p)/B^2$.  These are
therefore $O(\beta)$ corrections.
The grad-B and compression combination becomes a
curvature drift term appearing with $\ppp$, and its corrections, the
set now given by
\begin{equation}
  \div\vec J_G = \cdots 
  + \div\LB{\ppp\over B}\bunit\cross(\bdel\bunit)\RB
  - {\mu_0\over B^2}{1\over B}\grad\ppp\cdot\bunit\cross\grad\ppp
  + O(\beta)
\end{equation}
namely, the main curvature term with perpendicular pressure, to be
added to the one with parallel pressure, and corrections.
The second term is the main compressional correction, which now
cancels since 
\begin{equation}
  \grad\ppp\cdot\bunit\cross\grad\ppp = 0
\end{equation}
All other corrections are multiplied by $\mu_0\ppp/B^2$ or
$\mu_0\Delta p/B^2$, which are $O(\beta)$ small.

This is the main point.  The curvature correction term in the grad-B
drift 
due to the equilibrium
gradient, $\grad\ppp\equil$, adds to the compressional correction
term due to $B\grad\dbpl$, to form the combination just listed, which
cancels.  There are three main 
versions of this cancellation: linearised cases, nonlinear
fluctuations around a given equilibrium, or fully nonlinear cases with
a self consistent equilibrium.  We consider these in turn.

\subsection{linear cases}

In this case the cancellation is very similar to that found by Tang
\etal\ \cite{Tang80}.  The equilibrium magnetic
field includes the corrections given by finite current and pressure,
so that all changes are caused by the fluctuating dynamics.  The
curvature and grad-B drifts however differ by the correction due to
the equilibrium pressure gradient. 

There are two different terms, one due to the
grad-B drift, and one due to the compression.  In the grad-B drift,
the moment over $M\Omega$ gives just the fluctuation since
$\ppp\equil$ is given by the equilibrium correction.  So in
Eq.\ (\ref{eq:gradbpiece}), the factor of $\ppp$ just to the right of
the equals sign contains $\delta\ppp$ solely.
In the
compressional drift, since $\dbpl$ in Eq.\ (\ref{eq:gkcompress})
gives just the fluctuation
$\delta\ppp$, 
the moment over $M\Omega$ gives just the equilibrium.
So in Eq.\ (\ref{eq:deltabpiece}), the
factor of $\ppp$ just to the right of
the minus sign contains $\ppp\equil$ solely.
These compare as
\begin{equation}
  \grad(\delta\ppp) \cdot \bunit\cross\grad\ppp\equil
  + \grad\ppp\equil \cdot \bunit\cross\grad(\delta\ppp)
\end{equation}
respectively.  Exchanging $\ppp\equil$ and $\delta\ppp$ in one of the
terms produces the other term with a minus sign.
This is the cancellation between the two different
contributions as found in the 1980 reference just cited.

\subsection{nonlinear fluctuations on an equilibrium}

This is the version as just done.  The equilibrium magnetic
field includes the corrections given by finite current and pressure,
so that all changes are caused by the fluctuating dynamics.  The
curvature and grad-B drifts however differ by the correction due to
the equilibrium pressure gradient.  That much is the same as in a
linear case.  But in this one
the moment over $M\Omega$
gives the entire $\ppp$, equilibrium plus fluctuations, while the
correction in the grad-B drift term gives just the equilibrium piece
(Eq.\ \ref{eq:gradbpiece}),
and the compressional drift gives just the fluctuation
(Eq.\ \ref{eq:deltabpiece}).  These then
add into the entire $\ppp$, so that they combine into a single term
which vanishes identically.  A nonlinear delta-f case
with self consistent
gradients will go like this.  If the model treats fluctuation dynamics
only, with the equilibrium gradient present as an external drive, then
all terms except for quadratic nonlinearities are linearised and these
curvature and grad-B drift terms appear the same way as in a linear
model, as just described.

\subsection{nonlinear cases with self consistent equilibrium}

In a conventional tokamak, the parameter which defines the ordering is
$\eps^2/q^2\sim\beta\ll 1$, where
$\eps=a/R$, the inverse aspect ratio, and $q$ is the safety factor,
describing the toroidal/poloidal winding ratio of the magnetic field
lines on each toroidal flux surface \cite{HazeltineMeiss,Wesson}.
Together, $\eps/q$ gives the angle of the field line away from purely
toroidal.  It peaks at values close to $0.1$ or a little smaller in
conventional tokamaks.  One can consider taking just the vacuum
toroidal field, $I_0\grad\varphi$, with $I_0=B_0R_0$ constant, as the
guide field anchor for the gyrokinetic description.  In this case
\begin{equation}
  \bunit = R\grad\varphi \qquad \bdot\curl\bunit = 0
\end{equation}
as well as
\begin{equation}
  B = I_0/R \qquad \bdel\bunit = -\grad\log R = \grad\log B
\end{equation}
The result is that the curvature and grad-B drifts in the model are
identical, and $\dbpl$ responds to the entire perpendicular pressure.
So in Eq.\ (\ref{eq:gkcompress}), the factor of $\ppp$ on the right
contains the equilibrium piece, since it is fully self consistent.
The background equilibrium is the vacuum field, without any pressure.
In this case there is only the one correction term, namely, the one
due to the compression.  It is given by
\begin{equation}
  \vec J_G = \cdots
  + {\ppp\over B^2}\bunit\cross\grad(\dbpl)
\end{equation}
Putting in for $\dbpl$ using Eq.\ (\ref{eq:gkcompress}), we obtain
\begin{equation}
  \vec J_G = \cdots
  - {\mu_0\ppp\over B^2}\bunit\cross\grad{\ppp\over B}
\end{equation}
\begin{equation}
  \div\vec J_G = \cdots
  - {1\over B}{\mu_0\over B^2}\grad\ppp\cdot\bunit\cross\grad\ppp
  + O(\beta)
\end{equation}
which vanishes up to the $O(\beta)$ residuals.  
So the only real contributions due to the compression
enter at $O(\beta)$, not through $\grad\beta$, and they can be ordered
small.

In fact, 
conventional tokamak ordering can be
used 
in any of these scenarios 
in a conventional tokamak, and it is reasonable to set the above
approximations for $B$ and $\bunit$ in the Lagrangian, so that the
curvature and grad-B drifts are the same, and then $\aaperp$ hence
$\dbpl$ are neglected entirely in the Lagrangian.  This is in fact the
conventional approach.  It is only when one takes seriously a finite
beta equilibrium geometry as the background and/or finite beta
compressional responses, that this cancellation becomes important to
note, so that a seeming more accurate, but inconsistently flawed,
model leading to seemingly new effects does not become a temptation.

\subsection{Components of the Vector Potential in Compression}

A short accounting of how $\aaperp$ is the component of the total
vector potential involved in compression is included here for
completeness.  This is elementary information but may not be familiar
if one has only ever thought of MHD in terms of $\vec E$ and $\vec B$
rather than the potentials.

Given Faraday's law, we get the definition of $\vec E$, in terms of
the potentials $\vec A + \aapl\bunit + \aaperp$ and $\phi$, in this
case as
\begin{equation}
  \ptt{\vec B} = -\curl\vec E \qquad
  - \vec E = \ptt{\aapl}\bunit + \ptt{\aaperp} + \grad\phi
\end{equation}
since we take the background, including $\bunit$ as time independent.
The trivial version of this is to contract Faraday's law with
$\bunit$,
\begin{equation}
  \bdot\ptt{\vec B} = -\bdot\curl\vec E
  = \bunit\cdot\curl\ptt{\aaperp}
  = \ptt{}(\dbpl)
\end{equation}
with the pieces due to $\aapl$ and $\phi$ both vanishing.  The result
is not only trivial but also somewhat circular, since we used this
equation to define the potentials in the first place.

Slightly better is to consider the divergence of the ExB velocity,
\begin{equation}
  \div{1\over B}\bunit\cross\LP\grad\phi+\ptt{\vec a}\RP
  = \curl\LP{\bunit\over B}\RP\cdot\grad\phi
  + \div{\bunit\over B}\cross\ptt{\aaperp}
\end{equation}
where $\vec a=\aaperp+\bunit\aapl$ and $\bunit$ is static, belonging
to the background.  In a straight magnetic field ($\bb$ constant),
this reduces to
\begin{equation}
  \div{\exb\over B^2} = \div{\bunit\over B}\cross\ptt{\aaperp}
  = -{1\over B}\ptt{}(\dbpl)
  \label{eq:divexb}
\end{equation}
using Eq.\ (\ref{eq:dbpl}).  So $\dbpl$ hence $\aaperp$ is directly
related to compression of the general (not the electrostatic) ExB
velocity.

In MHD Faraday's law becomes the kinematic MHD equation.  If we
contract that with $\bunit/B$ and take the divergence of the MHD force
equation after dividing by $B^2$, we have
\begin{equation}
  {1\over B}\bunit\cdot\ptt{\vec B}
  = {1\over B}\ptt{}(\dbpl)
  = {1\over B}\bunit\cdot(\Bdel\vv-\vdel\bb)-\div\vv
  \label{eq:mhdcompress}
\end{equation}
\begin{equation}
  \div{\rhom\over B^2}\dtt{\vv} = \div(\bdel\bunit)
  - \div{\mu_0\over B^2}\grad p
  - \div{1\over B}\dpp\dbpl
  \label{eq:mhddiv}
\end{equation}
In the first equation $\aapl$ drops since $\bdot\bunit\cross()$
vanishes, and Eq.\ (\ref{eq:dbpl}) recasts $\curl\aaperp$.
In each case the last term is dominant if $\beta\ll 1$ and
the scale of motion is
much smaller than that of $\bunit$.  This shows that velocity
compression goes with $\dbpl$ in compression, and since $\dbpl$ is
given in terms of $\aaperp$, it is $\aaperp$ which is involved in
compression.  If the equation for $\aapl$ is derived, no velocity
divergence appears.  This has all been done in the development of
Reduced MHD in the years after it first appeared (see
esp.\ \cite{Park84,Park87}).

Hence the component that tracks magnetic compression due to
perpendicular divergence of velocity is solely $\aaperp$, including
situations with a quasistatic compression through a velocity
divergence, such as the divergence of the ExB velocity in an
inhomogeneous background magnetic field.  In the quasistatic case
Eq.\ (\ref{eq:gkcompress}) becomes the remnant of the MHD force
equation. 

\subsection{Whither the Fast Waves -- Magnetisation}

Although the low frequency limit precludes inclusion of fast
magnetosonic waves, the question of at which step they are ordered out
is of interest.  At low beta the wave velocity is just $v_A$, so the
frequency is $\kpp v_A$, which is very fast even in an electromagnetic
setting because $\kpp\gg\kpl$ by typically three orders of magnitude.
Indeed, we have $\kpp v_A=\Omega_i$ for $\kpp\rs=\beta_e^{1/2}$, where
$\beta_e = \mu_0p_e/B^2$ is the electron dynamical beta.  For tokamak
edge cases this point is near $\kpp\Lpp=1$, where $\Lpp$ is the
perpendicular profile scale length.  Additionally, at 
$\kpp\rs=1$ the frequency is $v_A/\rs=\beta_e^{-1/2}\Omega_i$, which
can be above $\wpi$, the ion plasma frequency, at which point even
quasineutrality would have to be dropped.

The first Lagrangian is the one considered in Eq.\ (\ref{eq:hhorig}).
In the Maxwell case the field Lagrangian going with this is the
one in Eq.\ (\ref{eq:maxwellfield}).  With the split into parallel and
perpendicular components in Eq.\ (\ref{eq:upl}), and with
quasineutrality, we simplify $\scriptl_f$ into
\begin{equation}
  \scriptl_f = -{1\over 2\mu_0}\abs{\dpp\aapl}^2
  -{1\over 2\mu_0}\abs{\curl\aaperp}^2
\end{equation}
The variation with respect to $\aaperp$ gives
\begin{equation}
  \curl\curl\aaperp = \mu_0 \intW fe\Uperp
\end{equation}
which is just the perpendicular component of Amp\`ere's law.
The variation of $L_p$ before invoking the smallness of $\drr$
(namely, Eq.\ \ref{eq:hhorig}) with respect to $\zpl$ and $\zperp$
gives 
\begin{equation}
  \dtt{\xx_\perp} = \Uperp \qquad \dtt{x_\parallel} = \Upl
\end{equation}
namely, simple velocities.  The variation with respect to $\xx$ gives
\begin{equation}
  e\dx\cross\Bstar - \dzperp - d\zpl\bunit = \grad H
\end{equation}
Note that $\Bstar$ includes $\bb$ and $\zpl\curl\bunit$ but nothing
from $\zperp$ (it does not appear multiplied by any
geometrical quantity).  Instead of solving for $\dx$ in terms of drifts and
parallel streaming, with $\zperp$ present we solve for it and $\zpl$
instead, to find
\begin{equation}
  \dtt{\zpl} = e\zpl(\curl\bunit)\cdot(\Uperp\cross\bunit)-\bdel H \qquad
  \dtt{\zperp} = \Uperp\cross\LP e\bb+\zpl\curl\bunit\RP-\dpp H
\end{equation}
This is very similar to the usual Lorentz force result, up to the
complications involving $\curl\bunit$ (the Lorentz force involving
$\aaperp$ is in the $d/dt$ on $\zpl$ and $\zperp$ and the part of $H$
involving $\aaperp$).  So there is no reason to expect the fast
magnetosonic waves to have dropped out at this point.

Things change at the point where we invoke the gyromotion.  After
processing that and evaluating the rest of $H$, we have a system in
which the square of $\zperp-e\aaperp-m\vexb$ divided by $B$ multiplies
$\dtheta$ and is therefore invariant.  Whatever the coordinate
description, the time derivative of this quantity remains zero.  So we
re-define it as the quantity $M$, note that it is an invariant giving
a version of the perpendicular energy, and write $L_p$ in terms of $M$
and $\vartheta$ instead of $\zperp$.  It is as this point that the
magnetosonic compressional waves are dropped from the system.  From
that point on, all perpendicular compression is quasistatic.

Under these conditions, the quasistatic force balance is just a
perturbed equilibrium, since this current comes from $\jxb=\grad p$
and the $\rho_G\vexb$ represents the nonlinear inertia correction,
clearly identified in this way with polarisation.
By contrast, the dynamics of the magnetosonic wave shows that it comes
from the reactive polarisation current (the $\ddt{\vv}$ in
Eq.\ \ref{eq:mhdforce}), as seen in
Eqs.\ (\ref{eq:mhdcompress},\ref{eq:mhddiv}) in which the main terms
are the time derivatives and the $\div\vv$ and $\dbpl$ terms on the
respective right hand sides.  This velocity divergence is one and the
same with the ExB divergence in Eq.\ (\ref{eq:divexb}).

The resolution of the issue is to note that the magnetisation
current is a vestige from the gyromotion (cf.\ pp.\ 36-7 of 
\cite{Freidberg})
The inertia piece of that
is one order down in this ordering scheme
(cf.\ Ref.\ \cite{Miyato15}).  In the Lie transform version of the
theory keeping all the high frequency components in the separate
equation for the gauge function in which all the gyrophase information
is stored, the fast magnetosonic waves are retained.  Indeed, a
central point of the 1999 work by H. Qin \etal, in
Ref.\ \cite{Qin99p}, was that ``The complete treatment of the
perpendicular current rendered by the gyrokinetic perpendicular
dynamics enables one to recover the compressional Alfv\'en wave from
the gyrokinetic model. From the viewpoint of gyrokinetic theory, the
physics of the compressional Alfv\'en wave is the polarization current
at second order. Therefore, in a low frequency gyrokinetic system, the
compressional Alfv\'en wave is naturally decoupled from the shear
Alfv\'en wave and drift wave'' (from the abstract).  This is the
central point of the position of the compressional waves in
gyrokinetic theory, and we recover the distinction in an heuristic way
in the above treatment.  When the conventional ordering is applied,
these waves are ordered with the high frequency dynamics, and the
resulting model keeps only the vestige compression resulting from the
quasistatic force balance evolving on the much slower time scales.  It
is helpful to have recovered this from a model more pedestrian
than the one in Ref.\ \cite{Qin99p}.

\section{Summary and Conclusion}

Gyrokinetic theory grew out of a method for the Hamiltonian basis for
mapping magnetic field lines \cite{CaryLittlejohn83}.  The variational
method for guiding center dynamics reaches back longer
\cite{Morozov66}.  Its development then merged with the Lie transform
techniques to form the gyrokinetic theory
\cite{Littlejohn81,Littlejohn82,Littlejohn83,
Dubin83,Littlejohn85,Hahm88,Hahm88a}.  
A review of the Lie transform methods and the general physics capture
by gyrokinetic theory was given in 2007 \cite{Brizard07}.

The original finding on the
compression cancellation -- that the difference due to the grad-B and
curvature drifts in the equilibrium was cancelled by the lowest order
low-beta correction due to compression -- came from an eikonal
formalism in an older style
for kinetic ballooning modes in toroidal plasmas, by Tang \etal\
in 1980 \cite{Tang80}.  This sat in parallel with the further development of
gyrokinetic theory, establishing the use of 
canonical form and the meaning of transformed magnetic coordinate $M$
\cite{Hahm88}, the 
Lie transform version of conversion of coordinates to $M$ or $\mu$ for
cases with finite flows
\cite{Brizard95,Hahm96}, the
role of high frequency forms and compression
\cite{Qin99p}, and its
application to MHD instabilities \cite{Naitou95} and
kinetic Alfv\'en waves \cite{Qin99,Lee01,Lee03},
the MHD equilibrium \cite{Qin00}.  Since about 2000, gyrokinetic
theory is a field theory supported by the theorems and method of
classical field theory, including the Noether theorem for conservation
laws \cite{Sugama00,Brizard00}.
Issues of momentum conservation, particularly the equivalence of
plasma and canonical toroidal anguler momentum in axisymmetric cases,
and the route back to the same results in MHD were clarified
\cite{momcons,Brizard11}.  The theory is also captured by the
Lie-Poisson functional bracket structure 
\cite{Morrison13,Brizard16}.

The main point of the present work has been to show that (1) the
finite beta cancellation result emerges naturally from gyrokinetic
theory, and that (2) it is covered not only for micro instabilities of
the ballooning type but also for general nonlinear dynamics as covered
by gyrokinetic and low frequency models of the reduced type
(perpendicular drifts rather than inertial flows or waves).  This
cancellation then holds for general, global computation which is
tractable at least for the conventional tokamak case.

The basic result of the cancellation is easy to understand: (1) the
difference between the grad-B and curvature drifts produces a
correction in the drifts proportional to $\bunit\cross\grad p\equil$,
(2) the compression $\dbpl$ produces an additional correction
proportional to $\bunit\cross\grad\delta p$, (3) both appear
multiplied by a factor of the pressure, (4) the combination is a drift
current proportional to $p\bunit\cross\grad p$, (5) the largest term
in the resulting divergence is $\grad p\cdot\bunit\cross\grad p$,
which vanishes, leaving small corrections all of which are $O(\beta)$,
and (6) all appearances of $p$ in these terms are equivalently $\ppp$, 
the perpendicular presssure in both the dynamics and the background
equilibrium.  

Concerning the compression, most of the connection between the
velocity divergence and variations in the strength of the magnetic
field is well known.  That the perpendicular velocity in MHD is just
the ExB one is also well known.  Faraday's law
and the description of the electric field in terms of both
electrostatic and magnetic potentials is also well known.  Put these
together and it follows directly that all of this MHD compression
physics occurs through the perpendicular components of the magnetic
potential, in the combination that produces the variations in the
field strength.  The connection with the polarisation current and the
gyrophase dependent part of the motion was made by Qin \etal\ in 1999
\cite{Qin99p}.  So when we make the low frequency approximations
around which conventional gyrokinetic theory is constructed, the part
of the compression that remains is the quasistatic force balance
associated with the largest piece in the magnetisation current.  And
if the plasma beta is small, then this effect can be safely neglected,
but solely if the grad-B drift is replaced by the curvature drift in
the current balance -- and hence the theory remains energetically
consistent if the relevant approximations are made in the system
Lagrangian upon which the model is based.

\begin{acknowledgments}
Fruitful discussions with F. Zonca and P. Lauber about the Tang
\etal\ (1980) version of this low-beta compression cancellation
are gratefully acknowledged.  F. Zonca made the recommendation noted
at the end of the Introduction to me at the APS/DPP 1999 meeting.
\end{acknowledgments}
{
\bibliography{../paper}
\bibliographystyle{aip}
}

\end{document}